\documentclass[preprint,3p,12pt,sort&compress]{elsarticle}

\makeatletter
\def\ps@pprintTitle{
    \let\@oddhead\@empty 
    \let\@evenhead\@empty
    \def\@oddfoot{\small \it{This manuscript is under review for publication in a peer-reviewed journal.}\hfil}
    \let\@evenfoot\@oddfoot
    }
\makeatother

\usepackage{lineno}
\usepackage{mathtools}
\usepackage{amssymb}
\usepackage{amsthm}
\usepackage{hyperref}
\usepackage{bm}
\usepackage{upgreek}
\usepackage{multirow}
\usepackage[nameinlink, capitalise, sort]{cleveref}
\usepackage{color}
\usepackage{colortbl}
\usepackage{float}
\usepackage{comment}
\usepackage{algorithm}
\usepackage{algpseudocode}
\hypersetup{pdfauthor=whatever}

\theoremstyle{definition}
\newtheorem{problem}{Problem}

\crefname{problem}{Problem}{Problems}
\Crefname{problem}{Problem}{Problems}

\crefname{thm}{Theorem}{Theorems}
\Crefname{thm}{Theorem}{Theorems}

\crefname{appendix}{}{}
\Crefname{appendix}{}{}

\begin{document}
\sloppy

\begin{frontmatter}
\title{Hierarchical Bayesian inversion using the Karhunen–Loève expansion with analytical eigenpairs of the squared exponential kernel}

\author[1]{Tatsuya Shibata}\corref{cor1}
\ead{shibata.tatsuya.84z@st.kyoto-u.ac.jp}
\author[2]{Michael C. Koch}
\author[1]{Kazunori Fujisawa}

\cortext[cor1]{Corresponding author}

\affiliation[1]{organization={Graduate School of Agriculture, Kyoto University},
addressline={Kitashirakawa Oiwake-cho},
city={Sakyo-ku},
postcode={606-8502 Kyoto},
country={Japan}}

\affiliation[2]{organization={Research School of Earth Sciences, Australian National University},
addressline={142 Mills Rd, Acton},
postcode={ACT 0200 Canberra},
country={Australia}}

\begin{abstract}
Hierarchical Bayesian inversion with Gaussian random field priors addresses uncertainty in covariance hyperparameters, such as the standard deviation and correlation length. 
When a Gaussian random field is represented by the Karhunen--Loève (KL) expansion, the basis functions depend on these hyperparameters through an integral eigenvalue problem (IEVP) associated with the covariance kernel. 
Consequently, the IEVP must be solved repeatedly whenever the hyperparameters are updated, leading to significant computational cost in hierarchical inference. 
In this paper, we focus on the squared exponential kernel and construct the KL expansion using the analytical solution to a Gaussian-weighted IEVP. 
This analytical KL expansion offers a computationally efficient alternative to the conventional KL expansion by eliminating the repeated numerical solutions of the IEVP during hyperparameter updates. 
While the analytical KL expansion is applicable to arbitrary domains and dimensions, it does not have the same mean-square optimality as the conventional KL expansion. To address this limitation, we employ an optimization-based approach that selects the standard deviation of the Gaussian weight function in the IEVP to effectively reduce the truncation error of the KL expansion.
Numerical experiments in one- and two-dimensional settings show that this selection strategy provides sufficient accuracy for practical applications.
Furthermore, the analytical KL expansion admits closed-form differentiation, enabling efficient posterior sampling via HMC.
The proposed framework is applied to Bayesian inversion for a steady Darcy flow model, where the hydraulic conductivity field is successfully estimated using weakly informative hyperpriors.
\end{abstract}


\end{frontmatter}

\section{Introduction}

In scientific and engineering fields such as geotechnical engineering, geophysics, and hydrogeology, inverse problems are widely studied. 
These problems involve inferring input information about a physical system from observational data. 
The unknown inputs are often represented as spatial fields (spatially varying functions), such as hydraulic conductivity, Young's modulus, or resistivity. 
Inverse problems are classically ill-posed \cite{Isakov2017,Kaipio2005}, which means that the existence and uniqueness of a solution, as well as its stability with respect to observational data, are not guaranteed. 
As a result, these problems are highly sensitive to uncertainties such as observation noise, making it difficult to obtain reliable solutions. 

To address these challenges, the Bayesian approach has become popular in recent years. 
In this framework, input information is modeled as a random field (a spatially varying random function), and the prior random field is updated using observational data based on Bayes' rule to obtain the posterior random field \cite{Kaipio2005,Stuart2010}. 
Introducing randomness allows for the derivation of not only single point estimates but also interval estimates, thereby enabling uncertainty quantification. 
Furthermore, a large class of practically relevant Bayesian inverse problems is well-posed under suitable notions of well-posedness introduced in \cite{Latz2023}. 
Thus, the Bayesian framework is a robust approach, providing both practical tools for uncertainty quantification and a theoretical guarantee of well-posedness.

Since posterior random fields are generally not available in closed-form, they are typically evaluated numerically using sampling methods such as Markov chain Monte Carlo (MCMC) \cite{Metropolis1953,Hastings1970,Gamerman2006}. 
To achieve this, the random field, which is an infinite-dimensional object, must be discretized into a finite number of random variables.
These random variables can then be treated as the unknown quantities in the inverse problem, and the posterior random field is reconstructed from their posterior distribution.

Series expansion methods represent random fields as infinite series with random coefficients, and discretization is achieved by truncating the series to a finite number of terms. 
Among these representations, the Karhunen--Loève (KL) expansion provides an optimal representation in the sense of minimizing the mean-square error for a given number of terms \cite{Ghanem1991}. 
This property makes it widely used for dimensionality reduction in inverse problems \cite{Latz2019,Marzouk2009,Sraj2016,Tagade2014,Siripatana2020,Polette2025,Shibata2025,Koch2020,Koch2021,Uribe2020}. 
In the KL expansion, the basis functions are constructed from the eigenvalues and eigenfunctions of the autocovariance operator that characterizes the random field. 
Consequently, solving a Fredholm integral eigenvalue problem (IEVP) is necessary, where the integral kernel is defined by the autocovariance function of the random field.

Gaussian random fields provide a simple modeling framework for spatially varying functions, characterized by a mean and a covariance structure \cite{Stuart2010}. 
Along with the selection of the mean and covariance function, hyperparameters such as the standard deviation and correlation length enable flexible control over the properties of the random field. 
Consequently, they are widely used as prior random fields. 
In practice, hyperparameters are rarely known a priori, and the choice of the prior can significantly influence the posterior distribution \cite{Latz2019}. 
Therefore, it is desirable to account for prior uncertainty through hierarchical Bayesian inference, where uncertain hyperparameters are treated as random variables.
For stationary Gaussian random fields represented by the KL expansion, the treatment of the mean and variance is relatively straightforward. 
This is because changes in the mean and variance correspond to translation and scaling, respectively, and the KL expansion can be efficiently updated through affine transformations of the KL basis \cite{Marzouk2009}. 
In contrast, the correlation length affects the KL basis and necessitates resolving the IEVP to obtain the eigenpairs at each step of MCMC \cite{Sraj2016}. 
To alleviate this computational burden, several surrogate modeling approaches have been proposed. 
In \cite{Tagade2014,Siripatana2020,Polette2025}, polynomial chaos (PC) expansion is employed. 
Approaches based on precomputed KL bases for multiple hyperparameter sets have been developed, including linear interpolation \cite{Mondal2014} and reduced basis (RB) methods \cite{Latz2019}.

An alternative approach to avoid recomputing the IEVP is to utilize analytical solutions. 
These solutions are available only for specific types of autocovariance functions (e.g., exponential \cite{Ghanem1991}, modified exponential \cite{Spanos2007}, and triangular \cite{Ghanem1991} kernels) defined on hyperrectangular domains. 
However, Mercer's theorem ensures that the domain on which the IEVP is solved does not need to coincide with the domain of the Gaussian random field (referred to as the physical domain). 
This property is referred to as the domain independence property of the KL expansion \cite{Pranesh2015}. 
In other words, for an arbitrarily shaped physical domain, a Gaussian random field can be represented by a KL expansion constructed from analytical solutions defined on the smallest hyperrectangular domain enclosing it (referred to as the bounding domain). 
It is important to note that such a KL expansion no longer minimizes the mean-square error over the physical domain. 
As a result, a larger number of terms is generally required to achieve accuracy comparable to that of the conventional KL expansion. 
Nevertheless, this increase is reported to be moderate \cite{Shibata2025,Papaioannou2012}, and the benefit of avoiding repeated IEVP solutions outweighs this drawback. 
The approximation error of the KL expansion based on the analytical solution of the exponential kernel
was investigated in \cite{Basmaji2023}.

In this paper, we adopt a similar approach and utilize an analytical solution to the IEVP of the squared exponential kernel \cite{Zhu1998,Rasmussen2005}. 
The squared exponential kernel yields infinitely differentiable sample paths; i.e., it is very smooth. Although Stein \cite{Stein1999} argues that such strong smoothness assumptions are unrealistic for modeling many physical processes, Rasmussen and Williams \cite{Rasmussen2005} note that this kernel is probably the most widely used in the kernel machines field. 
The squared exponential kernel is also commonly employed in Bayesian inversion, as demonstrated in previous studies \cite{Marzouk2009,Sraj2016,Tagade2014,Siripatana2020,Polette2025,Shibata2025,Koch2020,Koch2021,Uribe2020,Mondal2014,Iglesias2016}. 
One contributing factor is that it exhibits a rapid decay of eigenvalues \cite{Papaioannou2012,Huang2001,Sudret2000,Faes2022}, making it effective for dimensionality reduction through the KL expansion. 

However, the analytical solution is not employed in these studies, and its use for the KL basis appears to have received limited attention (e.g., \cite{Yin2023}).
This suggests that the analytical solution may have been regarded as unavailable for practical applications in the context of the KL expansion.
This view is also reflected in \cite{Faes2022}, where the lack of analytical solutions is noted. 
One possible reason is that the IEVP admitting an analytical solution is defined on $\mathbb{R}$ and involves a Gaussian-weighted integral, whereas the IEVP associated with the conventional KL expansion is typically defined on a bounded domain without such a weight. 
The KL basis derived from the weighted IEVP on $\mathbb{R}$ is orthogonal with respect to the weight over $\mathbb{R}$ and minimizes a weighted mean-square error over $\mathbb{R}$,
rather than the standard (unweighted) mean-square error over the physical domain. 
As a result, it does not have the classical optimality of the KL expansion.

Despite the loss of optimality, the analytical solution offers a significant advantage by eliminating the need to recompute the IEVP numerically in hierarchical Bayesian inversion.
Furthermore, the mean-square error can be made sufficiently small for practical purposes by appropriately choosing the standard deviation of the Gaussian weight function.
In this study, we propose an optimization-based approach for determining the standard deviation that minimizes the unweighted mean-square error. For computational simplicity, the error is evaluated over the hyperrectangle bounding domain rather than the physical domain.

It should also be noted that the validity of constructing the KL expansion via the weighted IEVP, together with its domain independence, is justified by Mercer's theorem, as described in \cite{Sun2005}. 
Additionally, higher-dimensional cases can be readily constructed using tensor products of the one-dimensional analytical solutions \cite{Basmaji2023,Fasshauer2012}. 
These properties enable the application of the KL expansion based on the analytical eigenpairs to arbitrary domains and dimensions.

Another important advantage is that the KL expansion based on the analytical eigenpairs can be differentiated in closed-form with respect to all parameters, including not only the random coefficients in the KL expansion but also the hyperparameters (i.e., the mean, standard deviation, and correlation length). 
This property enables efficient sampling using Hamiltonian Monte Carlo (HMC) \cite{Neal2011,Betancourt2018,Hoffman2014}, a gradient-based MCMC algorithm.
Furthermore, both the hyperparameters and the random coefficients in the KL expansion can be updated jointly at each HMC step without separating them.
This contrasts with Metropolis-within-Gibbs sampling \cite{Latz2019,Dunlop2017}, in which these components are updated sequentially.

The remainder of this paper is organized as follows. 
In \cref{sec:KLE}, we explain Mercer's theorem and then introduce the KL expansion. 
\cref{sec:SEK} presents the analytical solution to the Gaussian-weighted IEVP of the squared exponential kernel, including the selection of the standard deviation of the Gaussian weight function in the IEVP. 
In \cref{sec:VerifyAnalyticKL}, we numerically verify the accuracy of the KL expansion based on the analytical eigenpairs for one- and two-dimensional cases. \cref{sec:BayesianIP} explains the framework of hierarchical Bayesian inversion, including HMC. 
The gradient computation of the KL expansion for HMC is detailed in \cref{sec:GradComput}. 
Finally, \cref{sec:NumExp} presents numerical experiments of the inverse problem on hydraulic conductivity in two-dimensional steady Darcy flow.

\section{Karhunen--Loève expansion }\label{sec:KLE}

\subsection{Mercer's theorem}\label{subsec:Mercer's-theorem}

Let $X\subseteq\mathbb{R}^{d}$ denote the domain, and let $\nu$ be a non-degenerate Borel measure, i.e., $\nu(X')>0$ for any nonempty open set $X'\subseteq X$. 
Let $K:X\times X\rightarrow\mathbb{R}$ be a continuous, symmetric, and positive semi-definite kernel such that
\begin{equation}
\int_{X}\int_{X}\left|K(\mathbf{x},\mathbf{x}')\right|^{2}\mathrm{d}\nu(\mathbf{x})\,\mathrm{d}\nu(\mathbf{x}')<\infty.
\end{equation}
Then the associated integral operator $T_{K}$ on $L^{2}(X,\nu)$, defined by
\begin{equation}
(T_{K}f)(\mathbf{x})=\int_{X}K(\mathbf{x},\mathbf{x}')f(\mathbf{x}')\,\mathrm{d}\nu(\mathbf{x}'),
\end{equation}
is compact, self-adjoint, and positive. 
The operator $T_{K}$ has at most countably many positive eigenvalues $\left\{ \lambda_{i}\right\} ^{\infty}_{i=1}$, with corresponding eigenfunctions $\left\{ \phi_{i}\right\} ^{\infty}_{i=1}$ that are orthogonal in $L^{2}(X,\nu)$. 
For convenience, $\left\{ \lambda_{i}\right\} ^{\infty}_{i=1}$
are arranged in non-increasing order such that $\lambda_{i}\geq\lambda_{i+1}$
$(\lim_{i\rightarrow\infty}\lambda_{i}=0)$. 
Also, the eigenfunctions are chosen to be normalized; that is, $\int_{X}\phi_{i}(\mathbf{x})\phi_{j}(\mathbf{x})\,\mathrm{d}\nu(\mathbf{x})=\delta_{ij}$. 
The eigenpairs are obtained by solving the following integral eigenvalue problem (IEVP), which is a homogeneous Fredholm integral equation of the second kind:
\begin{equation}
\int_{X}K(\mathbf{x},\mathbf{x}')\phi_{i}(\mathbf{x}')\,\mathrm{d}\nu(\mathbf{x}')=\lambda_{i}\phi_{i}(\mathbf{x}).\label{eq:IEVP}
\end{equation}
Mercer's theorem on non-compact domains \cite{Sun2005} asserts that:
\begin{equation}
K(\mathbf{x},\mathbf{x}')=\sum^{\infty}_{i=1}\lambda_{i}\phi_{i}(\mathbf{x})\phi_{i}(\mathbf{x}'),\label{eq:Mercer}
\end{equation}
where the series converges absolutely and uniformly on $D\times D$ for any compact (i.e., closed and bounded) subsets $D\subset X$. 

\subsection{KL expansion of Gaussian random fields} \label{sec:kl_exp}
 
Let $\left(\Omega,\mathcal{F},P\right)$ be a probability space, and let $D\subset X$ be a bounded physical domain. Consider a real-valued Gaussian random field $u:D\times\Omega\rightarrow\mathbb{R}$ with a continuous mean $\mu:D\rightarrow\mathbb{R}$ and an autocovariance function $C:D\times D\rightarrow\mathbb{R}$,
that is, $u\sim\mathcal{GP}(\mu,C)$. 
The autocovariance function can be written as
\begin{equation}
C\left(\mathbf{x},\mathbf{x}'\right)=\sigma(\mathbf{x})\sigma(\mathbf{x}')\rho\left(\mathbf{x},\mathbf{x}'\right),
\end{equation}
where $\sigma:D\rightarrow\mathbb{R}$ is a standard deviation function, and $\rho:D\times D\rightarrow\mathbb{R}$ is an autocorrelation function.
Suppose that the autocorrelation function $\rho$ can be extended to a function on $X\times X$ that satisfies the same assumptions as the kernel $K$ in \cref{subsec:Mercer's-theorem}.
For stationary kernels defined by a single lag function, such as the squared exponential kernel, this extension is immediate since the kernel is already given on all of $\mathbb{R}^d\times\mathbb{R}^d$, where its positive semi-definiteness is guaranteed by Bochner's theorem \cite{Rasmussen2005}.
In the following, we identify this extended function with the kernel $K$. By Mercer's theorem, the eigendecomposition of $K$ in \cref{eq:Mercer}, when restricted to $D\times D$, converges absolutely and uniformly to the original autocorrelation function $\rho$. 
In other words, $\rho$, defined on $D\times D$, admits a series expansion with the eigenpairs $\left\{ \lambda_{i},\phi_{i}\right\} ^{\infty}_{i=1}$ of the IEVP
on $X$ (see \cref{eq:IEVP}). 
Therefore, the autocovariance function $C$ can be represented as
\begin{equation}
C\left(\mathbf{x},\mathbf{x}'\right)=\sigma(\mathbf{x})\sigma(\mathbf{x}')\sum^{\infty}_{i=1}\lambda_{i}\phi_{i}(\mathbf{x})\phi_{i}(\mathbf{x}'),\label{eq:Mercer2}
\end{equation}
and thus, the Gaussian random field $u$ defined on $D\times \Omega$ can be represented by the Karhunen--Loève expansion as
\begin{equation}
u(\mathbf{x},\omega)=\mu(\mathbf{x})+\sigma(\mathbf{x})\sum^{\infty}_{i=1}\sqrt{\lambda_{i}}\phi_{i}(\mathbf{x})\xi_{i}(\omega),\label{eq:KLE}
\end{equation}
where $\xi_{i}:\Omega\rightarrow\mathbb{R}$ are independent standard normal random variables. 
In fact, for any $\mathbf{x},\mathbf{x}'\in D$, the first- and second-order moments of $u(\mathbf{x},\omega)$ are given by
\begin{equation}
\begin{aligned}\mathrm{E}[u(\mathbf{x},\omega)] & =\mathrm{E}\left[\mu(\mathbf{x})+\sigma(\mathbf{x})\sum^{\infty}_{i=1}\sqrt{\lambda_{i}}\phi_{i}(\mathbf{x})\xi_{i}(\omega)\right]\\
 & =\mu(\mathbf{x})+\sigma(\mathbf{x})\sum^{\infty}_{i=1}\sqrt{\lambda_{i}}\phi_{i}(\mathbf{x})\mathrm{E}\left[\xi_{i}(\omega)\right]\\
 & =\mu(\mathbf{x})\quad(\text{since }\mathrm{E}\left[\xi_{i}(\omega)\right]=0),\label{eq:firstmoment}
\end{aligned}
\end{equation}
\begin{equation}
\begin{aligned}\mathrm{Cov}[u(\mathbf{x},\omega),u(\mathbf{x}',\omega)] & =\mathrm{E}\left[\sigma(\mathbf{x})\sigma(\mathbf{x}')\sum^{\infty}_{i=1}\sum^{\infty}_{j=1}\sqrt{\lambda_{i}}\phi_{i}(\mathbf{x})\xi_{i}(\omega)\sqrt{\lambda_{j}}\phi_{j}(\mathbf{x}')\xi_{j}(\omega)\right]\\
 & =\sigma(\mathbf{x})\sigma(\mathbf{x}')\sum^{\infty}_{i=1}\sum^{\infty}_{j=1}\sqrt{\lambda_{i}}\sqrt{\lambda_{j}}\phi_{i}(\mathbf{x})\phi_{j}(\mathbf{x}')\mathrm{E}\left[\xi_{i}(\omega)\xi_{j}(\omega)\right]\\
 & =\sigma(\mathbf{x})\sigma(\mathbf{x}')\sum^{\infty}_{i=1}\lambda_{i}\phi_{i}(\mathbf{x})\phi_{i}(\mathbf{x}')\quad(\text{since }\mathrm{E}\left[\xi_{i}(\omega)\xi_{j}(\omega)\right]=\delta_{ij})\\
 & =C(\mathbf{x},\mathbf{x}')\quad(\text{since }\mathrm{\cref{eq:Mercer2}}).\label{eq:secondmoment}
\end{aligned}
\end{equation}
This result indicates that the domain $X$ on which the IEVP is posed need not coincide with the physical domain $D$ on which the Gaussian random field is defined.
We refer to this as the domain independence property of the KL expansion, following \cite{Pranesh2015}. 
While \cite{Pranesh2015} demonstrated this property for bounded domains under the Lebesgue measure, \cref{eq:firstmoment,eq:secondmoment} show that it still holds in the more general setting of IEVPs posed on unbounded domains under any non-degenerate Borel measure.
Consequently, since Gaussian random fields are completely specified by their first- and second-order moments, samples of the Gaussian random field on $D$ can be generated using the KL expansion constructed with any $X\supseteq D$ and any such measure $\nu$.

In contrast, the accuracy of the approximation using the truncated KL expansion depends on the choice of $X$ and $\nu$. 
The truncated KL expansion is defined as
\begin{equation}
\hat{u}(\mathbf{x},\omega)=\mu(\mathbf{x})+\sigma(\mathbf{x})\sum^{M}_{i=1}\sqrt{\lambda_{i}}\phi_{i}(\mathbf{x})\xi_{i}(\omega).\label{eq:truncatedKL}
\end{equation}
Let $\varepsilon_{u}(\mathbf{x},\omega)=u(\mathbf{x},\omega)-\hat{u}(\mathbf{x},\omega)$ denote the approximation error.  
For a constant $\sigma$, the truncated KL expansion provides an optimal representation with respect to the expected $L^{2}(X,\nu)$-error, i.e., it minimizes
\begin{equation}
\mathrm{E}\left[\left\Vert \varepsilon_{u}(\cdot,\omega)\right\Vert ^{2}_{L^{2}(X,\nu)}\right]=\int_{X}\mathrm{E}\left[\varepsilon_{u}(\mathbf{x},\omega)^{2}\right]\mathrm{d}\nu(\mathbf{x}).\label{eq:weight_mse}
\end{equation}
This optimality essentially follows from the proof in \cite{Ghanem1991} of the KL expansion's optimality with respect to the Lebesgue measure; the main modification is the replacement of
$\mathrm{d}\mathbf{x}$ with $\mathrm{d}\nu(\mathbf{x})$.
In practice, however, the relevant measure of truncation error is the mean-square error over the physical domain $D$, measured with respect to the Lebesgue measure. 
This is referred to as the global mean-square error \cite{Betz2014}:
\begin{equation}
\overline{\varepsilon_{u}^2}=\mathrm{E}\left[\left\Vert \varepsilon_{u}(\cdot,\omega)\right\Vert ^{2}_{L^{2}(D)}\right]=\int_{D}\mathrm{E}\left[\varepsilon_{u}(\mathbf{x},\omega)^{2}\right]\mathrm{d}\mathbf{x}.\label{eq:mse}
\end{equation}
This error is minimized by the conventional KL expansion corresponding to the choice of $X=D$ and $\nu$ as the Lebesgue measure.

Although alternative choices of $X$ and $\nu$ do not minimize $\overline{\varepsilon_{u}^2}$, they can significantly facilitate the solution of the IEVP. 
In particular, by choosing $X$ as a hyperrectangular domain that encloses $D$, the IEVP in \cref{eq:IEVP} can be solved more easily. For certain kernels, such as the exponential \cite{Ghanem1991}, modified exponential \cite{Spanos2007}, and triangular \cite{Ghanem1991} kernels, analytical eigenpairs are available. 
Otherwise, the IEVP can be solved numerically using the Nyström method \cite{Betz2014} with a simple arrangement of integration points. 
Furthermore, the appropriate choice of $\nu$ enables the use of the analytical solution of the squared exponential kernel, as described in the next section.

In practical applications, since the domains and variance scales of random fields vary, it is useful to employ an appropriately normalized error metric for a unified treatment. 
The normalized variance of the truncation error, called the error variance \cite{Papaioannou2012,Sudret2000,Betz2014}, is given by
\begin{equation}
\varepsilon_{\sigma}(\mathbf{x};M)=\frac{\mathrm{Var}[u(\mathbf{x},\omega)-\hat{u}(\mathbf{x},\omega)]}{\mathrm{Var}[u(\mathbf{x},\omega)]}=1-\sum^{M}_{i=1}\lambda_{i}\phi_{i}(\mathbf{x})^{2}.
\end{equation}
A global error metric can then be obtained by averaging $\varepsilon_{\sigma}(\mathbf{x};M)$ over $D$, yielding the mean error variance \cite{Betz2014}, defined by:
\begin{equation}
\overline{\varepsilon_{\sigma}}(M)=\frac{1}{\left|D\right|}\int_{D}\varepsilon_{\sigma}(\mathbf{x};M)\,\mathrm{d}\mathbf{x}=1-
\frac{1}{\left|D\right|}\sum^{M}_{i=1}\lambda_{i}\int_{D}\phi_{i}(\mathbf{x})^{2}\,\mathrm{d}\mathbf{x}
,\label{eq:mev}
\end{equation}
where $\left|D\right|=\int_{D}\mathrm{d}\mathbf{x}$.
Note that, when the mean error variance is employed as the truncation error metric, it is natural to retain the $M$ KL terms with the largest values of $\lambda_{i}\int_{D}\phi_{i}(\mathbf{x})^{2}\,\mathrm{d}\mathbf{x}$.
For the conventional KL expansion, these $M$ terms can be selected solely based on the magnitude of $\lambda_{i}$ because $\int_{D}\phi_{i}(\mathbf{x})^2\,\mathrm{d}\mathbf{x}=1$. 
When using the domain independence property, however, the integral must be evaluated explicitly.
Hereafter, the dependence of $\varepsilon_{\sigma}$ and $\overline{\varepsilon_{\sigma}}$ on $M$ is omitted when it is clear from the context.

\section{Squared exponential kernel}\label{sec:SEK}

\subsection{Analytical solution}
Consider the one-dimensional squared exponential autocorrelation function defined by
\begin{equation}
\rho_{\mathrm{SE}}(x,x';l)=\exp\left(-\frac{\left|x-x'\right|^{2}}{l^{2}}\right),
\end{equation}
where $l$ is a correlation length. 
Let $X=\mathbb{R}$ and $\mathrm{d}\nu(x)=w(x;s)\,\mathrm{d}x$,
where $w(x;s)=\mathcal{N}(x|0,s^{2})$ is the Gaussian weight function with arbitrary $s>0$.
The analytical solution to \cref{eq:IEVP} for the squared exponential kernel is given as follows \cite{Zhu1998,Rasmussen2005}:
\begin{equation}
\lambda_{i}=\sqrt{\frac{2a}{A}}B^{i-1},\quad\phi_{i}(x)=\exp(-(c-a)x^{2})H_{i-1}(\sqrt{2c}x),\label{eq:eigpair}
\end{equation}
where $H_{i}$ is the $i$-th order Hermite polynomial, $a^{-1}=4s^{2}$, $b^{-1}=l^{2}$, and
\begin{equation}
c=\sqrt{a^{2}+2ab},\ A=a+b+c,\ B=b/A.
\end{equation}
For the application of the KL expansion, the eigenfunctions in \cref{eq:eigpair} must be normalized and then redefined as
\begin{equation}
\phi_{i}(x)=\left(\frac{c}{a}\pi\right)^{\frac{1}{4}}\exp(ax^{2})\psi_{i-1}\left(\sqrt{2c}x\right),
\end{equation}
where $\psi_{i}$ is the $i$-th Hermite function \cite{Walter1977} defined as
\begin{equation}
\psi_{i}(x)=\left(2^{i}i!\sqrt{\pi}\right)^{-\frac{1}{2}}\exp\left(-\frac{x^{2}}{2}\right)H_{i}(x).
\end{equation}
To clarify the dependence of the eigenpairs on $l$ and $s$, we rewrite $\lambda_i$ and $\phi_i$ as
\begin{equation}
\lambda_{i}(l,s)=\lambda\left(\gamma(l,s)\right)=\frac{2(\gamma-1)^{i-1}}{(\gamma+1)^{i}},\label{eq:ana_eigval}
\end{equation}
\begin{equation}
\phi_{i}(x;l,s)=\phi_{i}(x;\gamma(l,s),s)=(\gamma\pi)^{\frac{1}{4}}\exp\left(\frac{x^{2}}{4s^{2}}\right)\psi_{i-1}\left(\sqrt{\frac{\gamma}{2}}\frac{x}{s}\right),\label{eq:ana_eigfun}
\end{equation}
where $\gamma(l,s)=\sqrt{1+8s^{2}/l^{2}}$ is introduced for notational simplicity.

In the $d$-dimensional case, the squared exponential kernel is separable as
\begin{equation}
\rho_{\mathrm{SE}}(\mathbf{x},\mathbf{x}';\mathbf{l})=\prod^{d}_{n=1}\exp\left(-\frac{(x_{n}-x_{n}')^{2}}{l^{2}_{n}}\right),
\end{equation}
where $l_{n}$ is the correlation length for the $n$-th variable, $\mathbf{l}=(l_{1},\ldots,l_{d})$, and $\mathbf{x}=(x_{1},\ldots,x_{d})$.
Therefore, for $X=\mathbb{R}^{d}$ with $\mathrm{d}\nu(\mathbf{x})=w(\mathbf{x};\mathbf{s})\,\mathrm{d}\mathbf{x}$, where $w(\mathbf{x};\mathbf{s})=\mathcal{N}\left(\mathbf{x}|\mathbf{0},\mathrm{diag}(s^{2}_{1},\ldots,s^{2}_{d})\right)$, the eigenvalues and eigenfunctions can be readily obtained using the tensor product form \cite{Fasshauer2012}:
\begin{equation}
\lambda_{\bm{\upalpha}}(\mathbf{l},\mathbf{s})=\prod^{d}_{n=1}\lambda_{\alpha_{n}}(l_n,s_{n}),\quad\phi_{\bm{\upalpha}}(\mathbf{x};\mathbf{l},\mathbf{s})=\prod^{d}_{n=1}\phi_{\alpha_{n}}(x_{n};l_{n},s_{n}),\label{eq:multi_eigpair}
\end{equation}
where $\mathbf{s}=(s_{1},\ldots,s_{d})\in\mathbb{R}_+^d$, with $\mathbb{R}_+=(0,\infty)$,
$\bm{\upalpha}=(\alpha_{1},\ldots,\alpha_{d})\in\mathbb{N}^{d}$ is a multi-index, and $\left\{\lambda_{\alpha_{n}},\phi_{\alpha_{n}}\right\}$ denotes the $\alpha_{n}$-th eigenpair for the $n$-th variable. 
Note that when the weight function is centered at $\mathbf{m}\in\mathbb{R}^{d}$, i.e., $w(\mathbf{x};\mathbf{s})=\mathcal{N}\left(\mathbf{x}|\mathbf{m},\mathrm{diag}(s^{2}_{1},\ldots,s^{2}_{d})\right)$, the corresponding eigenfunctions are obtained by translating \cref{eq:multi_eigpair} by $\mathbf{m}$.

\subsection{Selection of the standard deviation of the Gaussian weight function}\label{subsec:select_s}

Consider the interpretation of the measure $\mathrm{d}\nu(\mathbf{x})=w(\mathbf{x};\mathbf{s})\,\mathrm{d}\mathbf{x}$. The Gaussian weight function $w(\mathbf{x};\mathbf{s})=\mathcal{N}\left(\mathbf{x}|\mathbf{m},\mathrm{diag}(s^{2}_{1},\ldots,s^{2}_{d})\right)$ decays rapidly away from $\mathbf{x}=\mathbf{m}$.
Therefore, the measure $\nu$ is effectively localized around $\mathbf{x}=\mathbf{m}$.
The standard deviation of the Gaussian weight function $\mathbf{s}$ determines the extent of this localization over the domain $X=\mathbb{R}^{d}$ \cite{Fasshauer2012}, which can be understood in terms of the mean-square error.
Throughout this subsection, the correlation length $\mathbf{l}$ is held fixed in order to focus on the effect of $\mathbf{s}$ on the truncation error of the KL expansion for a given covariance kernel. In this setting, the truncation error metrics inherit only the $\mathbf{s}$-dependence of the analytical eigenpairs. Hereafter, when this dependence needs emphasis, we explicitly include $\mathbf{s}$ as an additional argument of the corresponding error metric; otherwise, we omit it when the context is clear.
According to \cref{eq:weight_mse}, the KL expansion based on the analytical eigenpairs (hereafter referred to as the analytical KL expansion) minimizes the weighted mean-square error:
\begin{equation}
\int_{\mathbb{R}^{d}}\mathrm{E}\left[\varepsilon_{u}(\mathbf{x},\omega;\mathbf{s})^{2}\right]w(\mathbf{x};\mathbf{s})\,\mathrm{d}\mathbf{x}.
\end{equation}
Because $w(\mathbf{x};\mathbf{s})$ decays rapidly away from $\mathbf{x}=\mathbf{m}$, the approximation error $\varepsilon_{u}(\mathbf{x},\omega;\mathbf{s})$ in the tail regions contributes negligibly to this metric, allowing larger approximation errors in those regions.
Consequently, the contribution of the tail regions to the unweighted mean-square error $\overline{\varepsilon_{u}^2}(\mathbf{s})$ in \cref{eq:mse} may become large.
To maintain the accuracy of the KL expansion over $D$ with respect to $\overline{\varepsilon_{u}^2}(\mathbf{s})$, the parameter $\mathbf{s}$ should be chosen such that the significant mass of $w(\mathbf{x};\mathbf{s})$ is concentrated within $D$, while limiting the influence of its tails.
A more detailed mathematical interpretation of the choice of $\mathbf{s}$ can be found in \cite{Zhu1998,Fasshauer2012}.

However, in practice, even without a detailed understanding of such mathematical interpretation, the desired value of $\mathbf{s}$ can be obtained by solving one of the following optimization problems.
One is to determine the value of $\mathbf{s}$ that minimizes the mean error variance $\overline{\varepsilon_{\sigma}}(M,\mathbf{s})$ for a fixed number of KL terms $M$.
The other is, given a prescribed error tolerance, to first find the minimum $M$ that satisfies this tolerance, and
then find the value of $\mathbf{s}$ that minimizes $\overline{\varepsilon_{\sigma}}(M,\mathbf{s})$ for that $M$.
These problems can be summarized as follows.

\begin{problem}
\label{prob:1}For a truncated KL expansion with a fixed number of terms $M$, find $\mathbf{s}$ that minimizes the mean error variance
$\overline{\varepsilon_{\sigma}}(M,\mathbf{s})$; that is,
\begin{equation}
\mathbf{s}^{*}=\arg\min_{\mathbf{s}\in\mathbb{R}^{d}_{+}}\overline{\varepsilon_{\sigma}}(M,\mathbf{s}).
\end{equation}
\end{problem}

\begin{problem}
\label{prob:2}Given a prescribed tolerance $\overline{\varepsilon_{\sigma}}^{\mathrm{tol}}$ for the mean error variance $\overline{\varepsilon_{\sigma}}$, find the minimum number of KL terms $M$ that satisfies this tolerance;
that is,
\begin{equation}
M^{*}=\min\left\{
M\in\mathbb{N}\mid\exists\mathbf{s}\in\mathbb{R}^{d}_{+},\overline{\varepsilon_{\sigma}}(M,\mathbf{s})\leq\overline{\varepsilon_{\sigma}}^{\mathrm{tol}}\right\}.
\end{equation}
Then determine the corresponding optimal parameter $\mathbf{s}^{*}$ by solving \cref{prob:1} with $M=M^{*}$.
\end{problem}

For \cref{prob:2}, in practice, the primary objective is to obtain $M^{*}$ from the viewpoint of dimensionality reduction, while determining the corresponding $\mathbf{s}^{*}$ is optional. Therefore, any element in
$\left\{\mathbf{s}\in\mathbb{R}^{d}_{+}\mid\overline{\varepsilon_{\sigma}}(M^{*},\mathbf{s})\leq\overline{\varepsilon_{\sigma}}^{\mathrm{tol}}\right\}$
is also acceptable as a relaxed alternative.

The following approaches, for example, can be considered to solve these problems.
\cref{prob:1} for $d=1$ can be solved using either gradient-free optimization methods (e.g., grid search, Brent's method) or gradient-based optimization methods (e.g., Newton's method).
For $d\geq2$, the eigenvalues and eigenfunctions are constructed using tensor products of the one-dimensional analytical solutions, as shown in \cref{eq:multi_eigpair}.
In this case, when the KL terms are ordered by the magnitude of $\lambda_{\bm{\upalpha}}$ (or $\lambda_{\bm{\upalpha}}\int_{D}\phi_{\bm{\upalpha}}(\mathbf{x})^{2}\,\mathrm{d}\mathbf{x}$, as mentioned in \cref{sec:kl_exp}), their relative ordering may change between different multi-indices as $\mathbf{s}$ varies. This phenomenon is known as the crossing of eigenbranches \cite{Siripatana2020,Polette2025}. Consequently, the retained $M$ KL terms may change near such crossings, making the objective function $\overline{\varepsilon_{\sigma}}(M,\mathbf{s})$ locally non-differentiable with respect to $\mathbf{s}$.
Therefore, it is recommended to either use gradient-free optimization methods (e.g., grid search, particle swarm optimization) directly or, if employing gradient-based methods, to adopt a hybrid strategy that combines them with gradient-free optimization.
\cref{prob:2} can be addressed by exploiting the monotonicity $\min_{\mathbf{s}}\overline{\varepsilon_{\sigma}}(M+1,\mathbf{s})\leq\min_{\mathbf{s}}\overline{\varepsilon_{\sigma}}(M,\mathbf{s})$ and performing a binary search on $M$ to determine $M^{*}$.
At each step of the binary search, the same optimization as in \cref{prob:1} is conducted for the given $M$ to simultaneously determine the corresponding $\mathbf{s}$, as shown in \cref{alg:binary_search}.

\begin{algorithm}[t]
\caption{Solve \cref{prob:2} using the binary search}
\label{alg:binary_search}
\begin{algorithmic}
\Require Tolerance $\overline{\varepsilon_{\sigma}}^{\mathrm{tol}}$, Initial value $\mathbf{s}_{\mathrm{init}}$

\State $M_{\mathrm{low}}\gets0$
\State$M_{\mathrm{high}}\gets\min\left\{M\in\mathbb{N}\mid \overline{\varepsilon_{\sigma}}(M,\mathbf{s}_\mathrm{init})\le\overline{\varepsilon_{\sigma}}^{\mathrm{tol}}\right\}$
\State $\mathbf{s}^{*}_{M_{\mathrm{high}}}\gets\textsc{Solve Problem 1}(M_{\mathrm{high}})$ 

\While{$M_{\mathrm{high}}-M_{\mathrm{low}}>1$}
    \State $M_{\mathrm{mid}}\gets
    \left\lfloor
    (M_{\mathrm{low}}+M_{\mathrm{high}})/2
    \right\rfloor$
    \State $\mathbf{s}^{*}_{M_{\mathrm{mid}}}\gets\textsc{Solve Problem 1}(M_{\mathrm{mid}})$ 
    \If{$
    \overline{\varepsilon_{\sigma}}(M_{\mathrm{mid}},\mathbf{s}^{*}_{M_{\mathrm{mid}}})
    \le
    \overline{\varepsilon_{\sigma}}^{\mathrm{tol}}
    $}
        \State $M_{\mathrm{high}} \gets M_{\mathrm{mid}}$
        \State $\mathbf{s}^{*}_{M_\mathrm{high}} \gets \mathbf{s}^{*}_{M_\mathrm{mid}}$
    \Else
        \State $M_{\mathrm{low}} \gets M_{\mathrm{mid}}$
    \EndIf
    
\EndWhile
\State $M^{*}\gets M_{\mathrm{high}}$
\State $\mathbf{s}^{*}\gets \mathbf{s}^{*}_{M_{\mathrm{high}}}$
\State \Return $M^{*}$, $\mathbf{s}^{*}$
\end{algorithmic}
\end{algorithm}

Building on the above discussion, these optimization problems are quite complex (especially when $d\geq2$), and the choice of optimization strategies, as well as improvements in computational efficiency, may also be explored.
However, these aspects are beyond the scope of this study.
This is because, when the KL expansion is employed for dimensionality reduction of the inverse problem, these optimization problems need to be solved only once as a preprocessing step for inversion.
In other words, whichever optimization scheme is adopted, the associated computational cost is relatively small compared to that of the inversion as a whole.

In contrast, regardless of the optimization strategy, evaluating the mean error variance $\overline{\varepsilon_{\sigma}}(M,\mathbf{s})$ can become a bottleneck in implementation when $d\geq2$ due to the domain integration in \cref{eq:mev}.
For physical domains with complex geometries, designing quadrature rules is often cumbersome.
To alleviate this issue, one can introduce a bounding domain $D_{\mathrm{bound}}$, defined as the smallest hyperrectangular domain satisfying $D_{\mathrm{bound}}\supseteq D$ \cite{Pranesh2015}.
In \cref{prob:1,prob:2}, the original objective function $\overline{\varepsilon_{\sigma}}(M,\mathbf{s})$ can then be replaced with the mean error variance over this simpler domain:
\begin{equation}
\overline{\varepsilon_{\sigma}}^{\mathrm{bound}}(M,\mathbf{s})=\frac{1}{\left|D_{\mathrm{bound}}\right|}\int_{D_{\mathrm{bound}}}\varepsilon_{\sigma}(\mathbf{x};M,\mathbf{s})\,\mathrm{d}\mathbf{x}.\label{eq:mev_bound}
\end{equation}
This replacement facilitates the construction of quadrature rules and makes the implementation straightforward.
The practical validity of $M^{*}$ and $\mathbf{s}^{*}$ obtained from the substituted \cref{prob:2} by replacing $\overline{\varepsilon_{\sigma}}(M,\mathbf{s})$ with $\overline{\varepsilon_{\sigma}}^{\mathrm{bound}}(M,\mathbf{s})$ is examined in the next subsection.
It should be noted that, to efficiently localize the measure $\nu$ around $D_{\mathrm{bound}}$, it is natural to set $\mathbf{m}$ to be the center of $D_{\mathrm{bound}}$; we thus adopt this choice in this study.

\section{Verification of the analytical KL expansion via the mean error variance}\label{sec:VerifyAnalyticKL}

The accuracy of the analytical KL expansion is numerically investigated by comparing its mean error variance with that of the conventional KL expansion.
In the one-dimensional case, KL expansions are examined for multiple correlation lengths; in the two-dimensional case, for multiple physical domains with complex geometries. 
In constructing the analytical KL expansion, we consider only \cref{prob:2} since it subsumes \cref{prob:1}, and thus verification of \cref{prob:2} automatically entails verification of \cref{prob:1}. 
For all KL expansions, the mean is set to $\mu=\mathrm{const}$ and the standard deviation to $\sigma=1$, yielding $\overline{\varepsilon_{\sigma}}=\overline{\varepsilon_{u}^2}/|D|$.
Therefore, in these investigations, the conventional KL expansion with the Lebesgue measure provides the optimal approximation in terms of the mean error variance $\overline{\varepsilon_{\sigma}}$.

\subsection{One-dimensional case}

\begin{table}
\centering
\caption{Results of \cref{prob:2} for $D=[-1,1]$ with various combinations of $(l,\overline{\varepsilon_{\sigma}}^{\mathrm{tol}})$.}
\label{tab:1D_problem2}

\begin{tabular}{cccc}
\hline 
$l$ & $\overline{\varepsilon_{\sigma}}^{\mathrm{tol}}$ & $M^{*}$ & $s^{*}$\tabularnewline
\hline 
\multirow{3}{*}{$0.1$} & $10^{-2}$ & $28$ & $0.37626$\tabularnewline
                       & $10^{-3}$ & $38$ & $0.30723$\tabularnewline
                       & $10^{-4}$ & $46$ & $0.27256$\tabularnewline
\hline 
\multirow{3}{*}{$0.2$} & $10^{-2}$ & $14$ & $0.34769$\tabularnewline
                       & $10^{-3}$ & $19$ & $0.32493$\tabularnewline
                       & $10^{-4}$ & $23$ & $0.26172$\tabularnewline
\hline 
\multirow{3}{*}{$0.5$} & $10^{-2}$ & $6$  & $0.40712$\tabularnewline
                       & $10^{-3}$ & $8$  & $0.31757$\tabularnewline
                       & $10^{-4}$ & $10$ & $0.26001$\tabularnewline
\hline 
\multirow{3}{*}{$1$} & $10^{-2}$ & $4$ & $0.42819$\tabularnewline
                     & $10^{-3}$ & $5$ & $0.36348$\tabularnewline
                     & $10^{-4}$ & $6$ & $0.31596$\tabularnewline
\hline 
\end{tabular}
\end{table}

\begin{figure}[t]
\centering
\includegraphics{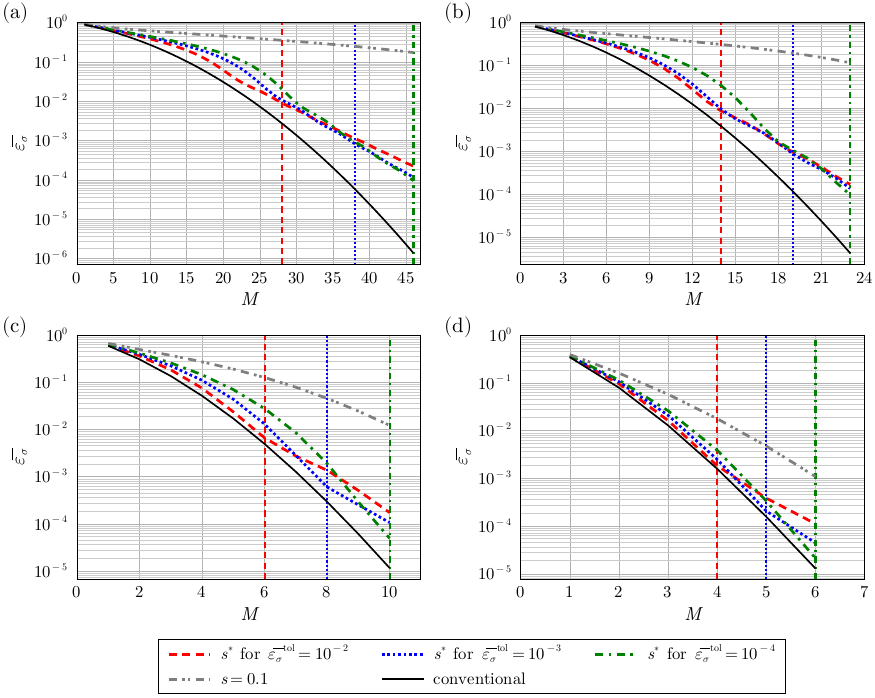}

\caption{Decay of the mean error variance for four analytical KL expansions and one conventional KL expansion for (a) $l=0.1$, (b) $l=0.2$, (c) $l=0.5$, and (d) $l=1$. 
In each panel, the three vertical lines indicate the positions of $M^{*}$ for each curve corresponding to the analytical KL expansion with $s^{*}$. 
Each vertical line matches the color and style of its respective curve.}
\label{fig:1D_mev}
\end{figure}

Consider the physical domain $D=[-1,1]$. 
For each correlation length $l=0.1,0.2,0.5,1$, \cref{prob:2} is solved using the tolerances $\overline{\varepsilon_{\sigma}}^{\mathrm{tol}}=10^{-2},10^{-3},10^{-4}$.
The resulting optimal parameters $M^{*}$ and $s^{*}$ are shown in \cref{tab:1D_problem2}. 
The conventional KL expansion is constructed using the Nyström method \cite{Betz2014} with 80 Gauss--Legendre quadrature points. 
These points are also used to evaluate the integral in the mean error variance $\overline{\varepsilon_{\sigma}}$ for all KL expansions.

\cref{fig:1D_mev} shows the decay of $\overline{\varepsilon_{\sigma}}$ for all KL expansions across four different correlation lengths. 
Each panel displays four curves for the analytical KL expansion: three corresponding to $s^{*}$ obtained for three tolerance levels, and one corresponding to $s=0.1$.
The position of $M^{*}$ for each curve corresponding to $s^{*}$ is indicated by a vertical line matching the curve's color and style. 
The decay rate of $\overline{\varepsilon_{\sigma}}$ depends on the choice of $s$. 
In particular, in all cases, the analytical KL expansion with the optimal parameter $s^{*}$ reaches the corresponding tolerance $\overline{\varepsilon_{\sigma}}^{\mathrm{tol}}$ faster than the other three analytical KL expansions. 
In contrast, the analytical KL expansion with $s=0.1$ decays significantly more slowly than these three analytical KL expansions. 
This highlights the importance of carefully selecting $s$. 

Additionally, as the correlation length $l$ or the tolerance $\overline{\varepsilon_{\sigma}}^{\mathrm{tol}}$ increase, the difference in the mean error variance between the analytical and the conventional KL expansion decreases. 
For example, for $(l,\overline{\varepsilon_{\sigma}}^{\mathrm{tol}})=(0.1,10^{-4})$, approximately 10 additional terms are required compared to the conventional
KL expansion even when using the optimal parameter $s^{*}$. 
In contrast, the difference diminishes for larger $l$ and $\overline{\varepsilon_{\sigma}}^{\mathrm{tol}}$; for example, for $(l,\overline{\varepsilon_{\sigma}}^{\mathrm{tol}})=(0.2,10^{-2})$ and $(l,\overline{\varepsilon_{\sigma}}^{\mathrm{tol}})=(0.5,10^{-4})$.

\subsection{Two-dimensional case}

\begin{figure}[t]
\centering
\includegraphics{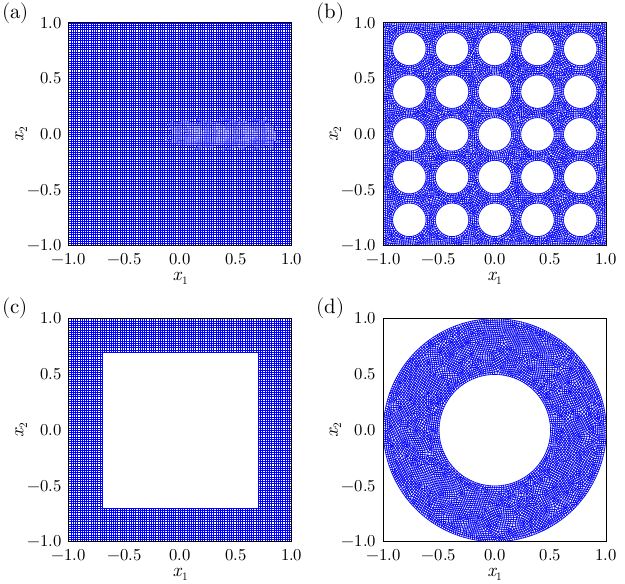}

\caption{Domains~A--D with their FE meshes: (a)--(d). Domain~A serves as the bounding domain for Domains~B--D.}
\label{fig:2D_test_domains}
\end{figure}

Consider Domains~A--D shown in \cref{fig:2D_test_domains}. 
Domain~A serves as the bounding domain $D_{\mathrm{bound}}=[-1,1]\times[-1,1]$, enclosing Domains~B--D, which have complex geometries. 
As described in \cref{subsec:select_s}, \cref{prob:2} is addressed on $D_{\mathrm{bound}}$ rather than on each physical domain to simplify implementation. 
For correlation lengths $l_{1}=l_{2}=0.5$, \cref{prob:2} is solved on $D_{\mathrm{bound}}$ using the tolerances $\overline{\varepsilon_{\sigma}}^{\mathrm{tol}}=10^{-2},10^{-3},10^{-4}$, where the integral in $\overline{\varepsilon_{\sigma}}^{\mathrm{bound}}$ (see \cref{eq:mev_bound}) is evaluated using $20\times20$ Gauss--Legendre quadrature points. 
The resulting optimal parameters $M^{*}$ and $\mathbf{s}^{*}$ are shown in \cref{tab:2D_problem2} and are commonly used for the analytical KL expansion for Domains B--D. 
The conventional KL expansion is constructed using the Galerkin method \cite{Papaioannou2012,Sudret2000,Betz2014} with the finite element (FE) meshes shown in \cref{fig:2D_test_domains}.
These meshes are also used to evaluate the integral in the mean error variance $\overline{\varepsilon_{\sigma}}$ for all KL expansions.

\begin{table}
\centering
\caption{Results of \cref{prob:2} for $D_{\mathrm{bound}}=[-1,1]\times[-1,1]$ and $l_{1}=l_{2}=0.5$ with several tolerance levels.}
\label{tab:2D_problem2}

\begin{tabular}{ccc}
\hline 
$\overline{\varepsilon_{\sigma}}^{\mathrm{tol}}$ & $M^{*}$ & $\mathbf{s}^{*}$\tabularnewline
\hline 
$10^{-2}$ & $38$ & $(0.39914,0.39379)$\tabularnewline
$10^{-3}$ & $63$ & $(0.31953,0.31953)$\tabularnewline
$10^{-4}$ & $94$ & $(0.32567,0.32215)$\tabularnewline
\hline 
\end{tabular}
\end{table}

\begin{figure}
\centering
\includegraphics{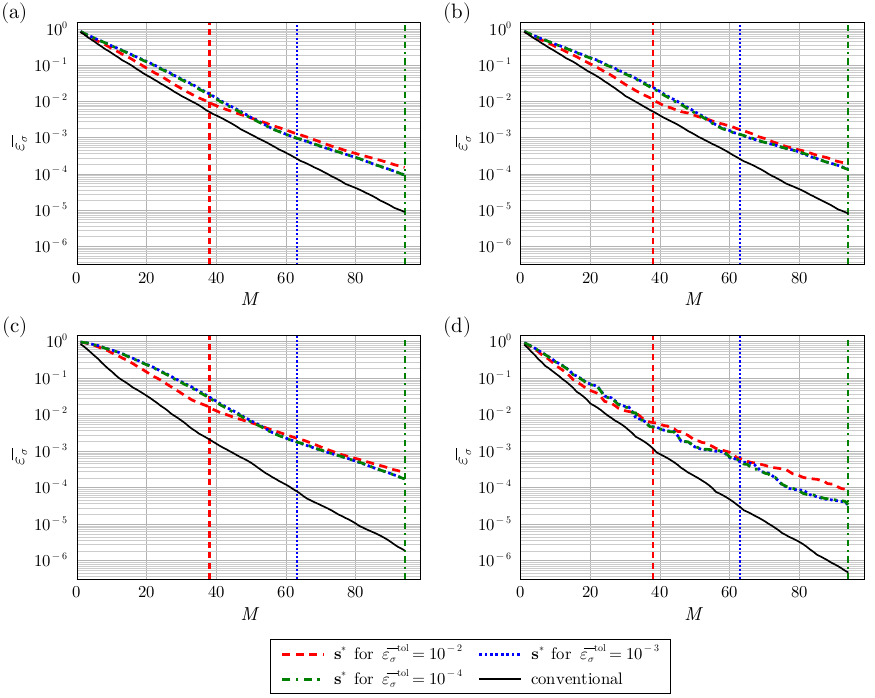}

\caption{Decay of the mean error variance of the KL expansions for $l_{1}=l_{2}=0.5$.
(a)--(d) correspond to Domains~A--D, respectively. For each tolerance
level $\overline{\varepsilon_{\sigma}}^{\mathrm{tol}}$, the optimal
parameter $s^{*}$ for the bounding domain $D_{\mathrm{bound}}$ is
consistently used across all analytical KL expansions. The position
of $M^{*}$ is indicated by a vertical line with the same color and
style as the corresponding curve.}\label{fig:2D_mev}
\end{figure}

\begin{figure}
\centering
\includegraphics{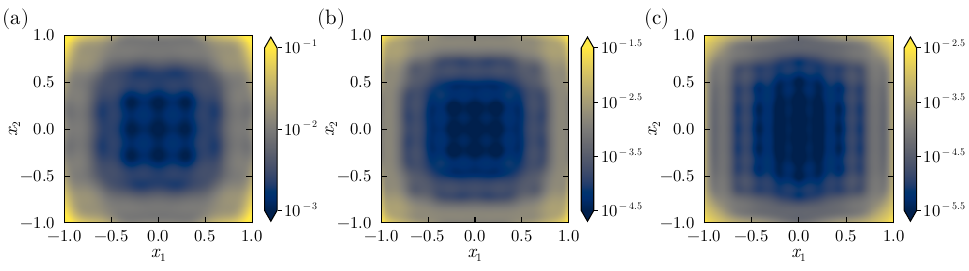}

\caption{The error variance $\varepsilon_{\sigma}(\mathbf{x})$ of the analytical KL expansion on $D_{\mathrm{bound}}$ with the optimal parameters $s^{*}$ and $M^{*}$ for (a) $\overline{\varepsilon_{\sigma}}^{\mathrm{tol}}=10^{-2}$, (b) $\overline{\varepsilon_{\sigma}}^{\mathrm{tol}}=10^{-3}$, and
(c) $\overline{\varepsilon_{\sigma}}^{\mathrm{tol}}=10^{-4}$.}
\label{fig:2D_ev}
\end{figure}

\cref{fig:2D_mev} shows the decay of $\overline{\varepsilon_{\sigma}}$ for all KL expansions on Domains~A--D. 
For Domains~B~and~C, $\overline{\varepsilon_{\sigma}}$ of the analytical KL expansion does not satisfy the tolerance level $\overline{\varepsilon_{\sigma}}^{\mathrm{tol}}$. However, it is slightly larger than that for Domain~A, and this difference can be considered negligible in practice.
For Domain~D, $\overline{\varepsilon_{\sigma}}$ of the analytical KL expansion is smaller than that for Domain~A and satisfies the tolerance level $\overline{\varepsilon_{\sigma}}^{\mathrm{tol}}$.
This behavior of $\overline{\varepsilon_{\sigma}}$ can also be confirmed from \cref{fig:2D_ev}, which shows the error variance $\varepsilon_{\sigma}(\mathbf{x})$ on $D_{\mathrm{bound}}$ using the optimal parameters for each tolerance $\overline{\varepsilon_{\sigma}}^{\mathrm{tol}}$. 
In all cases, $\varepsilon_{\sigma}(\mathbf{x})$ tends to increase with distance from the origin. 
Consequently, Domains~B~and~C exclude the region near the origin where $\varepsilon_{\sigma}(\mathbf{x})$ is small, leading to a higher $\overline{\varepsilon_{\sigma}}$ than that for Domain~A. 
In contrast, Domain~D excludes the outer regions where $\varepsilon_{\sigma}(\mathbf{x})$ is large, resulting in a lower $\overline{\varepsilon_{\sigma}}$ than that for Domain~A.
These results indicate that, unless the physical domain $D$ is an extreme shape consisting only of regions where $\varepsilon_{\sigma}(\mathbf{x})$ is large, $M^{*}$ and $\mathbf{s}^{*}$ optimized for $D_{\mathrm{bound}}$ achieve accuracy on $D$ comparable to that on $D_{\mathrm{bound}}$.
Therefore, they are practically sufficient in most cases.

However, two points should be noted. 
First, $M^{*}$ and $\mathbf{s}^{*}$ optimized for $D_{\mathrm{bound}}$ are not in general optimal for $D$. 
This can be confirmed from \cref{fig:2D_mev}(d), where $\mathbf{s}^{*}$ for $\overline{\varepsilon_{\sigma}}^{\mathrm{tol}}=10^{-2}$ does not minimize the mean error variance $\overline{\varepsilon_{\sigma}}$ at the corresponding $M^*$ (i.e., $M=38$).
Instead, $\mathbf{s}^{*}$ for $\overline{\varepsilon_{\sigma}}^{\mathrm{tol}}=10^{-3}$ and $10^{-4}$ yield lower $\overline{\varepsilon_{\sigma}}$.
Second, a lower $\overline{\varepsilon_{\sigma}}$ for the analytical KL expansion does not imply that the expansion is close to optimal.
As mentioned earlier, $\overline{\varepsilon_{\sigma}}$ for the analytical KL expansion is smallest for Domain~D among Domains~A--D.
However, its difference from the corresponding conventional KL expansion, i.e., the optimal representation, is not minimal. 
Specifically, the difference for Domain~D is larger than those for Domains~A~and~B, as shown in \cref{fig:2D_mev}. 

\section{Inverse problem}\label{sec:BayesianIP}

\subsection{Bayesian inference}

Consider the Gaussian random field $u\sim\mathcal{GP}(\mu,C_{\mathrm{SE}})$, where
\begin{equation}
C_{\mathrm{SE}}(\mathbf{x},\mathbf{x}';\sigma,\mathbf{l})=\sigma^{2}\rho_{\mathrm{SE}}(\mathbf{x},\mathbf{x}';\mathbf{l})
\end{equation}
is the squared exponential autocovariance function. 
Hereafter, the mean $\mu$ and the standard deviation $\sigma$ are constant on the physical domain $D$, that is, $\mu=\mu(\mathbf{x})$ and $\sigma=\sigma(\mathbf{x})$.
In this setting, $u$ can be parameterized by the random coefficients $\bm{\upxi}=(\xi_{1},\ldots,\xi_{M})\in\mathbb{R}^{M}$ in the truncated KL expansion (see \cref{eq:truncatedKL}), along with hyperparameters $\bm{\uptau}=(\mathbf{l},\sigma,\mu)\in\mathbb{R}^{d+2}$. 
For simplicity, these parameters are combined into $\bm{\uptheta}=(\bm{\upxi},\bm{\uptau})\in\Theta\subseteq\mathbb{R}^{M+d+2}$, where $\Theta$ is the parameter space.
Then $u$ can be expressed in the following form, explicitly indicating its dependence on parameters:
\begin{equation}
u(\mathbf{x};\bm{\uptheta})=\mu+\sigma\sum^{M}_{i=1}\sqrt{\lambda_{i}(\mathbf{l})}\phi_{i}(\mathbf{x};\mathbf{l})\xi_{i}.\label{eq:truncatedKL2}
\end{equation}
Here, the eigenpairs $\lambda_{i}$ and $\phi_{i}$ also depend on the standard deviation of the Gaussian weight function $\mathbf{s}$. 
However, since $\mathbf{s}$ is fixed at its optimal value $\mathbf{s}^{*}$ during inversion, this dependence is omitted for clarity, and the eigenpairs are denoted simply by $\lambda_{i}(\mathbf{l})$ and $\phi_{i}(\mathbf{x};\mathbf{l})$, respectively. 
To ensure positivity, we adopt a log-normal random field $k$ as a prior, given by
\begin{equation}
k(\mathbf{x};\bm{\uptheta})=10^{u(\mathbf{x};\bm{\uptheta})}.
\end{equation}

Consider the forward problem of evaluating $G(k_{\bm{\uptheta}})$, where $G:V\rightarrow W$ is a forward operator between Banach spaces $V$ and $W$, and $k_{\bm{\uptheta}}=k(\cdot;\bm{\uptheta})\in V$ represents a realization of the random field. 
Let $O:W\rightarrow Y$ be an observation operator that maps $G(k_{\bm{\uptheta}})\in W$ to noise-free observations, where $Y\subseteq\mathbb{R}^{N}$ is the data space.
Accordingly, the parameter-to-observation operator $\mathcal{G}:\Theta\rightarrow Y$ is defined by
\begin{equation}
\mathcal{G}(\bm{\uptheta})=(O\circ G)(k(\cdot;\bm{\uptheta})).\label{eq:G2G}
\end{equation}
In practice, the measured observations are contaminated by noise and can be modeled as
\begin{equation}
\mathbf{y}=\mathcal{G}(\bm{\uptheta})+\bm{\upeta},\label{eq:obs_model}
\end{equation}
where $\mathbf{y}\in Y$ denotes the observations, and $\bm{\upeta}\in\mathbb{R}^{N}$ is observation noise. 
This noise is typically assumed to follow a zero-mean Gaussian distribution $\mathcal{N}(\mathbf{0},\mathbf{R})$, where $\mathbf{R}\in\mathbb{R}^{N\times N}$ is a covariance matrix.
In inverse problems, the unknown input $k_{\bm{\uptheta}}$ is recovered from $\mathbf{y}$ by estimating $\bm{\uptheta}=(\bm{\upxi},\bm{\uptau})$.
Typically, from the perspective of computational cost, only $\bm{\upxi}$ is estimated while $\bm{\uptau}$ is fixed, either based on apriori information or determined through maximum likelihood estimation \cite{Rasmussen2005,Tipireddy2020}.
This is because the IEVP in \cref{eq:IEVP} is generally solved numerically, and repeatedly evaluating the IEVP every time $\bm{\uptau}$ (essentially $\mathbf{l}$) is updated during the inversion would incur prohibitive computational costs.
In contrast, this study employs the analytical solution to the IEVP, which makes the computational cost negligible.
Consequently, $\bm{\uptau}$ can also be treated as a random parameter to be estimated within the hierarchical Bayesian framework.

In the Bayesian approach, a prior distribution of the estimated parameters $\bm{\uptheta}$ is assumed as initial knowledge.
This distribution is updated with observations $\mathbf{y}$ to obtain the posterior distribution, which is the solution of the inverse problem. 
Let the prior distribution be $p(\bm{\uptheta})=p(\bm{\upxi})p(\bm{\uptau})$, i.e., $\bm{\upxi}$ and $\bm{\uptau}$ are independent.
The posterior distribution $p(\bm{\uptheta}|\mathbf{y})$ is then obtained using Bayes' rule:
\begin{equation}
p(\bm{\uptheta}|\mathbf{y})\propto p(\mathbf{y}|\bm{\uptheta})p(\bm{\uptheta})=p(\mathbf{y}|\bm{\uptheta})p(\bm{\upxi})p(\bm{\uptau}),\label{eq:Bayes_rule}
\end{equation}
where $p(\mathbf{y}|\bm{\uptheta})$ is the likelihood. Considering \cref{eq:obs_model}, the likelihood is given by
\begin{equation}
p(\mathbf{y}|\bm{\uptheta})=\mathcal{N}(\mathbf{y}|\mathcal{G}(\bm{\uptheta}),\mathbf{R})\propto\exp\left(-\frac{1}{2}(\mathbf{y}-\mathcal{G}(\bm{\uptheta}))^{\top}\mathbf{R}^{-1}(\mathbf{y}-\mathcal{G}(\bm{\uptheta}))\right).\label{eq:likelihood}
\end{equation}
It should be noted that although $k$ and $\bm{\uptau}$ have a hierarchical structure, the KL expansion allows $k$ to be parameterized by $\bm{\upxi}$, which is independent of $\bm{\uptau}$. 
This formulation is referred to as the non-centered parameterization \cite{Betancourt2015}. 
As a result, the parameters explicitly used in the estimation are $\bm{\upxi}$ and $\bm{\uptau}$, and \cref{eq:Bayes_rule} does not include a probability density associated with the conditional random field $k|$$\bm{\uptau}$. 

In many practical problems, the posterior distribution $p(\bm{\uptheta}|\mathbf{y})$ cannot be expressed in closed-form. 
This is because the forward model $\mathcal{G}(\bm{\uptheta})$ in the likelihood $p(\mathbf{y}|\bm{\uptheta})$ is often given in an implicit form with respect to $\bm{\uptheta}$, as is typical in problems involving partial differential equations
(PDEs). 
Therefore, $p(\bm{\uptheta}|\mathbf{y})$ is numerically obtained by generating samples via Markov chain Monte Carlo (MCMC) methods. In MCMC algorithms, the constant term of the target distribution cancels out when calculating the acceptance probability.
Therefore, up to a normalizing constant, $p(\bm{\uptheta}|\mathbf{y})$ can be written as a canonical distribution:
\begin{equation}
p(\bm{\uptheta}|\mathbf{y})\propto\exp(-U(\bm{\uptheta})),\label{eq:posterior}
\end{equation}
where $U(\bm{\uptheta})$ denotes the negative log-density. Expressing the prior as $p(\bm{\uptheta})\propto\exp(-E(\bm{\uptheta}))$, Bayes' rule in \cref{eq:Bayes_rule}, along with the likelihood in \cref{eq:likelihood}, yields
\begin{equation}
U(\bm{\uptheta})=\frac{1}{2}(\mathbf{y}-\mathcal{G}(\bm{\uptheta}))^{\top}\mathbf{R}^{-1}(\mathbf{y}-\mathcal{G}(\bm{\uptheta}))+E(\bm{\uptheta}).\label{eq:potential_energy}
\end{equation}

\subsection{Hamiltonian Monte Carlo}

Hamiltonian Monte Carlo (HMC) is a gradient-based MCMC algorithm that uses Hamiltonian dynamics to efficiently explore the parameter space \cite{Neal2011,Betancourt2018,Hoffman2014}.
In HMC, an auxiliary momentum variable $\mathbf{p}$, having the same dimension as $\bm{\uptheta}$, is introduced. 
In the standard implementation, $\mathbf{p}$ is drawn independently of $\bm{\uptheta}$ from a multivariate
normal distribution $p(\mathbf{p})=\mathcal{N}(\mathbf{p}|\mathbf{0},\mathbf{M})$,
resulting in the canonical joint distribution:
\begin{equation}
p(\bm{\uptheta},\mathbf{p}|\mathbf{y})=p(\mathbf{p})p(\bm{\uptheta}|\mathbf{y})\propto\exp(-H(\mathbf{p},\bm{\uptheta})).
\end{equation}
Here, $H(\mathbf{p},\bm{\uptheta})$ is referred as to the Hamiltonian and is defined by
\begin{equation}
H(\mathbf{p},\bm{\uptheta})=U(\bm{\uptheta})+\frac{1}{2}\mathbf{p}^{\top}\mathbf{M}^{-1}\mathbf{p}.
\end{equation}
In this context, $U(\bm{\uptheta})$ is the potential energy, $(\mathbf{p}^{\top}\mathbf{M}^{-1}\mathbf{p})/2$
is the kinetic energy, and $\mathbf{M}$ is called the mass matrix.
The introduction of the Hamiltonian enables the deterministic trajectories in the sampling process through Hamiltonian dynamics:
\begin{equation}
\frac{\mathrm{d}\bm{\uptheta}}{\mathrm{d}t}=\frac{\partial H}{\partial\mathbf{p}}(\mathbf{p},\bm{\uptheta})=\mathbf{M}^{-1}\mathbf{p},
\end{equation}
\begin{equation}
\frac{\mathrm{d}\mathbf{p}}{\mathrm{d}t}=-\frac{\partial H}{\partial\bm{\uptheta}}(\mathbf{p},\bm{\uptheta})=-\frac{\partial U}{\partial\bm{\uptheta}}(\bm{\uptheta}).
\end{equation}
These equations are usual solved numerically using the leapfrog method, which satisfies time reversibility and volume conservation properties. 
The time discretization of the leapfrog method consists of three steps:
\begin{equation}
\mathbf{p}\left(t+\frac{\epsilon}{2}\right)=\mathbf{p}\left(t\right)-\frac{\epsilon}{2}\frac{\partial U}{\partial\bm{\uptheta}}(\bm{\uptheta}(t)),
\end{equation}
\begin{equation}
\bm{\uptheta}\left(t+\epsilon\right)=\bm{\uptheta}\left(t\right)+\epsilon\mathbf{M}^{-1}\mathbf{p}\left(t+\frac{\epsilon}{2}\right),
\end{equation}
\begin{equation}
\mathbf{p}\left(t+\epsilon\right)=\mathbf{p}\left(t+\frac{\epsilon}{2}\right)-\frac{\epsilon}{2}\frac{\partial U}{\partial\bm{\uptheta}}(\bm{\uptheta}(t+\epsilon)).
\end{equation}
Starting from the $j$-th state $(\bm{\uptheta}_{j},\mathbf{p}_{j})$, the $(j+1)$-th state $(\mathbf{\bm{\uptheta}}_{j+1},\mathbf{p}_{j+1})$ is generated through two steps that leave the canonical joint distribution
invariant. 
In the first step, $\mathbf{p}_{j}$ is replaced by a new random draw from $\mathcal{N}(\mathbf{0},\mathbf{M})$. 
In the second step, the $(j+1)$-th sample candidate $(\bm{\uptheta}',\mathbf{p}')$ is proposed from the current state $(\mathbf{\bm{\uptheta}}_{j},\mathbf{p}_{j})$ by applying the leapfrog steps $L$ times with step-size $\epsilon$.
The proposed state is then accepted or rejected as $(\mathbf{\bm{\uptheta}}_{j+1},\mathbf{p}_{j+1})$ according to the Metropolis-Hastings acceptance criterion \cite{Afshar2021}:
\begin{equation}
\begin{aligned}\alpha((\mathbf{\bm{\uptheta}}_{j},\mathbf{p}_{j}),(\mathbf{\bm{\uptheta}}',\mathbf{p}')) & =\min\left\{ 1,\frac{p(\mathbf{\bm{\uptheta}}',\mathbf{p}')}{p(\bm{\uptheta}_{j},\mathbf{p}_{j})}\left|\frac{\partial(\bm{\uptheta}',\mathbf{p}')}{\partial(\bm{\uptheta}_{j},\mathbf{p}_{j})}\right|\right\} \\
 & =\min\left\{ 1,\exp(-H(\bm{\uptheta}',\mathbf{p}')+H(\bm{\uptheta}_{j},\mathbf{p}_{j}))\right\} .
\end{aligned}
\end{equation}
Here, $\left|\partial(\mathbf{\bm{\uptheta}}',\mathbf{p}')/\partial(\mathbf{\bm{\uptheta}}_{j},\mathbf{p}_{j})\right|$ is the Jacobian for the conversion from $(\bm{\uptheta}_{j},\mathbf{p}_{j})$ to $(\bm{\uptheta}',\mathbf{p}')$, and $\left|\partial(\mathbf{\bm{\uptheta}}',\mathbf{p}')/\partial(\mathbf{\bm{\uptheta}}_{j},\mathbf{p}_{j})\right|=1$ due to volume conservation. 
The choice of $\epsilon$, $L$, and $\mathbf{M}$ in the leapfrog integrator affects the sampling efficiency of HMC. 
This study employs automatic tuning methods for these parameters, with an adaptation procedure similar to Stan \cite{Carpenter2017}.
In detail, the Dual Averaging scheme \cite{Nesterov2009} is used for tuning $\epsilon$, the No-U-Turn Sampler \cite{Hoffman2014} is used for tuning $L$, and the adaptation of $\mathbf{M}$ is done according to section 4.2.1 in \cite{Betancourt2018}.

In HMC implementation, some libraries (e.g., AdvancedHMC.jl \cite{Xu2020} in Julia or HamiltonianSampler in MATLAB) allow users to avoid implementing the entire HMC algorithm from scratch; it is sufficient to specify the target distribution and its gradient. Thus, one only needs to define $U(\bm{\uptheta})$ and $\partial U(\bm{\uptheta})/\partial\bm{\uptheta}$. 

\section{Gradient computation in HMC}\label{sec:GradComput}

\subsection{Gradient of the potential energy}

The gradient of the potential energy with respect to $\bm{\uptheta}$ is obtained by differentiating \cref{eq:potential_energy}:
\begin{equation}
\frac{\partial U}{\partial\bm{\uptheta}}(\bm{\uptheta})=
-(\mathbf{y}-\mathcal{G}(\bm{\uptheta}))^{\top}\mathbf{R}^{-1}\frac{\partial\mathcal{G}}{\partial\bm{\uptheta}}(\bm{\uptheta})+\frac{\partial E}{\partial\bm{\uptheta}}(\bm{\uptheta}),\label{eq:grad_potential_energy}
\end{equation}
Considering \cref{eq:G2G}, the $j$-th component of $\partial\mathcal{G}(\bm{\uptheta})/\partial\bm{\uptheta}$ is given by
\begin{equation}
\frac{\partial\mathcal{G}}{\partial\theta_{j}}(\bm{\uptheta})=D(O\circ G)(k_{\bm{\uptheta}})\left[\frac{\partial k_{\bm{\uptheta}}}{\partial\theta_{j}}\right],\label{eq:dg_dtheta}
\end{equation}
where $D(O\circ G)(k_{\bm{\uptheta}})$ denotes the Fréchet derivative of $O\circ G$ at $k_{\bm{\uptheta}}$, and
\begin{equation}
\frac{\partial k_{\bm{\uptheta}}}{\partial\theta_{j}}=k_{\bm{\uptheta}}\ln10\frac{\partial u}{\partial\theta_{j}}(\cdot;\bm{\uptheta}).
\end{equation}
Here, $\partial u/\partial\theta_{j}$ is the partial derivative of the KL expansion, which can be derived analytically as detailed in \cref{subsec:GradKL}.
By applying the chain rule, \cref{eq:dg_dtheta} can be expressed as
\begin{equation}
\frac{\partial\mathcal{G}}{\partial\theta_{j}}(\bm{\uptheta})=DO(G(k_{\bm{\uptheta}}))\left[DG(k_{\bm{\uptheta}})\left[\frac{\partial k_{\bm{\uptheta}}}{\partial\theta_{j}}\right]\right]=DO(G(k_{\bm{\uptheta}}))\left[\frac{\partial G(k_{\bm{\uptheta}})}{\partial\theta_{j}}\right],\label{eq:dg_dtheta2}
\end{equation}
where $\partial G(k_{\bm{\uptheta}})/\partial\theta_{j}$ is the sensitivity of the forward problem with respect to $\theta_{j}$. 
Evaluating $\partial G(k_{\bm{\uptheta}})/\partial\theta_{j}$
can be computationally expensive when the forward problem is governed by a PDE, making it prohibitive to compute these sensitivities for all parameters. 
To address this issue, the adjoint method provides an effective strategy \cite{Koch2020,Fichtner2006,Aghasi2012}.
By introducing an adjoint field, this method eliminates the need to explicitly compute the sensitivities $\partial G(k_{\bm{\uptheta}})/\partial\theta_{j}$ when evaluating
$\partial U(\bm{\uptheta})/\partial\bm{\uptheta}$. Consequently, $\partial U(\bm{\uptheta})/\partial\bm{\uptheta}$ can be computed with only a single adjoint solve following the forward solve, significantly reducing the computational cost. 
For details, see \cref{app:adjoint_method}.

\subsection{Gradient computation of the KL expansion}\label{subsec:GradKL}

One advantage of the analytical KL expansion is that its partial derivatives can be expressed in closed-form. 
Theses partial derivatives for each parameter (i.e., $\bm{\upxi}$, $\mathbf{l}$, $\sigma$, and $\mu$) are obtained by differentiating \cref{eq:truncatedKL2}:
\begin{equation}
\frac{\partial u}{\partial\xi_{i}}(\mathbf{x};\bm{\uptheta})=\sigma\sqrt{\lambda_{i}(\mathbf{l})}\phi_{i}(\mathbf{x};\mathbf{l})\quad(i=1,\ldots,M),
\end{equation}
\begin{equation}
\frac{\partial u}{\partial l_{m}}(\mathbf{x};\bm{\uptheta})=\sigma\sum^{M}_{i=1}\left(\frac{1}{2\sqrt{\lambda_{i}(\mathbf{l})}}\frac{\partial\lambda_{i}}{\partial l_{m}}(\mathbf{l})\cdot\phi_{i}(\mathbf{x};\mathbf{l})+\sqrt{\lambda_{i}(\mathbf{l})}\frac{\partial\phi_{i}}{\partial l_{m}}(\mathbf{x};\mathbf{l})\right)\xi_{i}\quad(m=1,\ldots,d),
\end{equation}
\begin{equation}
\frac{\partial u}{\partial\sigma}(\mathbf{x};\bm{\uptheta})=\sum^{M}_{i=1}\sqrt{\lambda_{i}(\mathbf{l})}\phi_{i}(\mathbf{x};\mathbf{l})\xi_{i},
\end{equation}
\begin{equation}
\frac{\partial u}{\partial\mu}(\mathbf{x};\bm{\uptheta})=1.
\end{equation}
Here, $\partial\lambda_{i}(\mathbf{l})/\partial l_{m}$ and $\partial\phi_{i}(\mathbf{x};\mathbf{l})/\partial l_{m}$ correspond to the partial derivatives of \cref{eq:multi_eigpair}.
Let $\bm{\upalpha}_{i}$ be the multi-index corresponding to the $i$-th eigenpair. 
Then these derivatives are given by
\begin{equation}
\frac{\partial\lambda_{i}}{\partial l_{m}}(\mathbf{l})=\frac{\partial\lambda_{\alpha_{i,m}}}{\partial l_{m}}(l_{m})\prod^{d}_{\substack{n=1\\
n\neq m
}
}\lambda_{\alpha_{i,n}}(l_{n}),
\end{equation}
\begin{equation}
\frac{\partial\phi_{i}}{\partial l_{m}}(\mathbf{x};\mathbf{l})=\frac{\partial\phi_{\alpha_{i,m}}}{\partial l_{m}}(x_{m};l_{m})\prod^{d}_{\substack{n=1\\
n\neq m
}
}\phi_{\alpha_{i,n}}(x_{n};l_{n}).
\end{equation}
Hereafter, the derivatives of the one-dimensional eigenpairs are considered.
For notational simplicity, $\partial\lambda_{i}(l)/\partial l$ and $\partial\phi_{i}(x;l)/\partial l$ are used in place of $\partial\lambda_{\alpha_{i,m}}(l_{m})/\partial l_{m}$ and $\partial\phi_{\alpha_{i,m}}(x_{m};l_{m})/\partial l_{m}$ throughout in this section. 
By the chain rule, these derivatives are expressed as
\begin{equation}
\frac{\partial\lambda_{i}}{\partial l}(l)=\frac{\partial}{\partial l}\lambda_{i}\left(\gamma(l;s^{*})\right)=\frac{\partial\lambda_{i}}{\partial\gamma}\left(\gamma(l;s^{*})\right)\cdot\frac{\partial\gamma}{\partial l}(l;s^{*}),
\end{equation}
\begin{equation}
\frac{\partial\phi_{i}}{\partial l}(x;l)=\frac{\partial}{\partial l}\phi_{i}(x;\gamma(l;s^{*}),s^{*})=\frac{\partial\phi_{i}}{\partial\gamma}(x;\gamma(l;s^{*}),s^{*})\cdot\frac{\partial\gamma}{\partial l}(l;s^{*}),
\end{equation}
where
\begin{equation}
\frac{\partial\gamma}{\partial l}(l;s^{*})=-\frac{8s^{*}{}^{2}}{l^{3}\sqrt{1+\dfrac{8s^{*}{}^{2}}{l^{2}}}}.
\end{equation}
The partial derivatives of eigenpairs with respect to $\gamma$ are obtained by differentiating \cref{eq:ana_eigval,eq:ana_eigfun}:
\begin{equation}
\frac{\partial\lambda_{i}}{\partial\gamma}(\gamma)=(4i-2\gamma-2)\frac{(\gamma-1)^{i-2}}{(\gamma+1)^{i+1}},
\end{equation}
\begin{equation}
\frac{\partial\phi_{i}}{\partial\gamma}(x;\gamma,s^{*})=\frac{1}{4}\left(\frac{\pi}{\gamma^{3}}\right)^{\frac{1}{4}}\exp\left(\frac{x^{2}}{4s^{*}{}^{2}}\right)\left(\psi_{i-1}\left(\sqrt{\frac{\gamma}{2}}\frac{x}{s^{*}}\right)+\frac{\sqrt{2\gamma}x}{s^{*}}\psi'_{i-1}\left(\sqrt{\frac{\gamma}{2}}\frac{x}{s^{*}}\right)\right),
\end{equation}
where the derivative of the Hermite function $\psi'_{i}$ is given by \cite{Walter1977}
\begin{equation}
\psi'_{i}(x)=\sqrt{\frac{i}{2}}\psi_{i-1}(x)-\sqrt{\frac{i+1}{2}}\psi_{i+1}(x).
\end{equation}

\section{Numerical experiments of the inverse problem}\label{sec:NumExp}

We consider an inverse problem using the steady Darcy flow model, governed by Darcy's law and the continuity equation:
\begin{equation}
\mathbf{q}=-k\nabla h,\quad\nabla\cdot\mathbf{q}=Q\quad\text{in }D,\label{eq:darcy_continuity}
\end{equation}
where $k$ is the hydraulic conductivity, $h$ is the total hydraulic head, $\mathbf{q}$ is the volumetric flux, and $Q$ is the source term. 
Combining these two equations yields the following PDE:
\begin{equation}
-\nabla\cdot\left(k\nabla h\right)=Q\quad\text{in }D.\label{eq:PDE}
\end{equation}
Assume that the boundary $\partial D$ consists of the Dirichlet boundary $\Gamma_{\mathrm{D}}$ and the Neumann boundary $\Gamma_{\mathrm{N}}$ (i.e., $\partial D=\Gamma_{\mathrm{D}}\cup\Gamma_{\mathrm{N}}$, $\Gamma_{\mathrm{D}}\cap\Gamma_{\mathrm{N}}=\emptyset$), with the boundary conditions given by
\begin{equation}
h=h_{\mathrm{D}}\quad\mathrm{\text{on }}\Gamma_{\mathrm{D}},\quad\mathbf{q}\cdot\mathbf{n}=q_{\mathrm{N}}\quad\text{on }\Gamma_{\mathrm{N}}.\label{eq:bc}
\end{equation}
Here, $h_{\mathrm{D}}$ and $q_{\mathrm{N}}$ are given boundary data, and $\mathbf{n}$ denotes the outward unit normal vector on the boundary.
The unknown hydraulic conductivity $k$ is estimated using two types of observations: the hydraulic head and the flow rate. Let $O_{h}\coloneqq\left(O_{h,1},\ldots,O_{h,N_{h}}\right)$ and $O_{q}\coloneqq\left(O_{q,1},\ldots,O_{q,N_{q}}\right)$ be observation operators for the hydraulic head and the flow rate, respectively.
Here and in the following, the subscripts $h$ and $q$ indicate association with the hydraulic head and the flow rate. 
Each component of $O_{h}$ and $O_{q}$ corresponds to an individual observation. 
Specifically, the hydraulic head at the observation point $\mathbf{x}_{n}\in D$
$\left(n=1,\ldots,N_{h}\right)$ is given by
\begin{equation}
O_{h,n}(h)=\int_{D}\delta(\mathbf{x}-\mathbf{x}_{n})h(\mathbf{x})\,\mathrm{d}\mathbf{x}.\label{eq:obs_h}
\end{equation}
Similarly, the flow rate over the boundary segments $\Gamma_{q,n}\ (n=1,\ldots,N_{q})$, which are mutually disjoint (i.e., $\Gamma_{q,n}\cap\Gamma_{q,m}=\emptyset$ for $n\neq m$), is given by
\begin{equation}
O_{q,n}(\mathbf{q})=\int_{\Gamma_{q,n}}\mathbf{q}(\mathbf{x})\cdot\mathbf{n}\,\mathrm{d}S(\mathbf{x}).\label{eq:obs_q}
\end{equation}
Note that the flow rate is observed on the Dirichlet boundary $\Gamma_{\mathrm{D}}$, since the flow rate on the Neumann boundary $\Gamma_{\mathrm{N}}$ is prescribed and provides no additional information. 
Hence, $\Gamma_{q,n}\subset\Gamma_{\mathrm{D}}$ for $n=1,\ldots,N_{q}$.

Let $G_{h}$ be the forward operator for $h$ (i.e., $h=G_{h}(k)$), and $G_{q}$ be the forward operator for $\mathbf{q}$ (i.e., $\mathbf{q}=G_{q}(k)$).
The corresponding parameter-to-observation operators are then defined as $\mathcal{G}_{h}(\bm{\uptheta})=\left(O_{h}\circ G_{h}\right)(k(\cdot;\bm{\uptheta}))$
and $\mathcal{G}_{q}(\bm{\uptheta})=\left(O_{q}\circ G_{q}\right)(k(\cdot;\bm{\uptheta}))$, respectively. 
Assuming mutually independent Gaussian observation noise, the observation model in \cref{eq:obs_model} can be written as
\begin{equation}
\left[\begin{array}{c}
\mathbf{y}_{h}\\
\mathbf{y}_{q}
\end{array}\right]=\left[\begin{array}{c}
\mathcal{G}_{h}(\bm{\uptheta})\\
\mathcal{G}_{q}(\bm{\uptheta})
\end{array}\right]+\left[\begin{array}{c}
\bm{\upeta}_{h}\\
\bm{\upeta}_{q}
\end{array}\right],\quad\left[\begin{array}{c}
\bm{\upeta}_{h}\\
\bm{\upeta}_{q}
\end{array}\right]\sim\mathcal{N}\left(\left[\begin{array}{c}
\mathbf{0}\\
\mathbf{0}
\end{array}\right],\left[\begin{array}{cc}
\mathbf{R}_{h} & \mathbf{0}\\
\mathbf{0} & \mathbf{R}_{q}
\end{array}\right]\right),\label{eq:gauss_obs_model}
\end{equation}
where $\mathbf{R}_{h}$ and $\mathbf{R}_{q}$ denote diagonal covariance matrices corresponding to the observation noises $\bm{\upeta}_{h}$ and $\bm{\upeta}_{q}$, respectively. 
It is worth mentioning that, regardless of the choice of prior for $\bm{\uptheta}$, the Bayesian inverse problem associated with the additive Gaussian noise model in \cref{eq:gauss_obs_model} is well-posed in at least the weak, Hellinger, and total variation senses defined in \cite{Latz2023}.

We consider two inverse problems for the steady Darcy flow model.
Test 1 is similar to the benchmark problem used in \cite{Iglesias2016,Dunlop2017,Iglesias2013}. 
Test 2 is defined on a complex geometric domain and aims to demonstrate the effectiveness of the analytical KL expansion for arbitrary domains.
Throughout this section, all physical quantities are expressed in SI units. 
The units of key physical quantities are listed in \cref{tab:SI}.
Hereafter, units are omitted unless otherwise stated.

\begin{table}
\centering
\caption{Units of key physical quantities. Descriptions of these quantities can be found in \cref{sec:NumExp}.}
\label{tab:SI}

\begin{tabular}{cc}
\hline 
Quantity & Unit\tabularnewline
\hline 
$x_{1},x_{2},l_{1},l_{2},h,O_{h,i}(h)$ & $\mathrm{m}$\tabularnewline
$k$ & $\mathrm{m}/\mathrm{s}$\tabularnewline
$O_{q,i}(\mathbf{q})$ & $\mathrm{m}^{2}/\mathrm{s}$\tabularnewline
$u,\sigma,\mu$ & --\tabularnewline
\hline 
\end{tabular}
\end{table}

\subsection{Test 1 }\label{subsec:Test1}

\begin{figure}[t]
\centering
\includegraphics{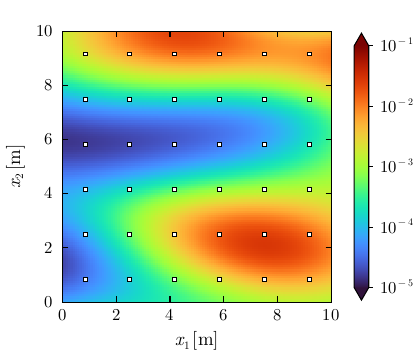}

\caption{The true hydraulic conductivity field generated through the log-hydraulic conductivity $u$ represented by the KL expansion with $(l_{1},l_{2},\sigma,\mu)=(6,3,1,-3)$.
White squares indicate observation points for the hydraulic head.}\label{fig:test1_config}
\end{figure}

Consider the hydraulic conductivity field $k$ on $D=[0,10]\times[0,10]$ as shown in \cref{fig:test1_config}. 
The true field is generated numerically through the log-hydraulic conductivity $u$, represented by the KL expansion with hyperparameters $(l_{1},l_{2},\sigma,\mu)=(6,3,1,-3)$.
The source term is $Q=0$, and the boundary conditions are defined as 
\begin{equation}
\begin{aligned}h(x_{1},0)=0, & \quad\frac{\partial h}{\partial x_{1}}(10,x_{2})=0,\\
-k\frac{\partial h}{\partial x_{1}}(0,x_{2})=5\times10^{-4}, & \quad\frac{\partial h}{\partial x_{2}}(x_{1},10)=0.
\end{aligned}
\end{equation}
These conditions are similar to the benchmark setup used in \cite{Iglesias2016,Dunlop2017,Iglesias2013}.
The observational data include only the total hydraulic head at $36$ points equally distributed over $D$ (white squares in \cref{fig:test1_config}).
The noise-free observations are synthetically generated using FEM with $120\times120$ FE meshes, and then contaminated by independent noise with a standard deviation of $10\%$ of their respective values.
The prior for $k$ is set to the log-normal random field; that is, the prior for $u$ is given by $u\sim\mathcal{GP}(\mu,C_{\mathrm{SE}}(\mathbf{\cdot},\cdot;\sigma,\mathbf{l}))$.
The prior for the random coefficients $\bm{\upxi}$ in the KL expansion is the standard normal $\bm{\upxi}\sim\mathcal{N}(\mathbf{0},\mathbf{I})$, while hyperparameters are assigned weakly informative priors: $\log_{10}(l_{1}/l^{\min}_{1})\sim\mathcal{HN}(0,1)$,
$\log_{10}(l_{2}/l^{\min}_{2})\sim\mathcal{HN}(0,1)$, $\sigma\sim\mathcal{HN}(0,1)$,
and $\mu\sim\mathcal{N}(-4,2)$. 
The priors for $\sigma$ and $\mu$ are selected such that the induced Gaussian random fields adequately represent hydraulic conductivity relevant to engineering applications, ranging approximately from $10^{-8}$ to $10^{0}\:\mathrm{m}/\mathrm{s}$ \cite{Terzaghi1996}. 
The priors for the correlation lengths $l_{n}$ $(n=1,2)$ ensure $l_{n}\geq l^{\min}_{n}$, while allowing $l_{n}$ to take
values up to roughly $10^{2}l^{\min}_{n}$. 
The lower bounds are set to $l^{\min}_{1}=l^{\min}_{2}=1$, which corresponds to one-tenth of the domain length in each spatial direction. 
This choice excludes excessively short correlation lengths while still retaining sufficient flexibility for the random field to represent fine-scale spatial variability.
The prescribed tolerance of the mean error variance is set to $\overline{\varepsilon_{\sigma}}=10^{-2}$, and the number of terms in the KL expansion is chosen such that this tolerance is satisfied even at $l_{1}=l^{\min}_{1}$ and $l_{2}=l^{\min}_{2}$.
To this end, the number of terms $M$ is set to $M^{*}=222$, which is obtained by solving \cref{prob:2} under the conditions $\overline{\varepsilon_{\sigma}}=10^{-2}$, $l_{1}=l^{\min}_{1}$, $l_{2}=l^{\min}_{2}$, and $D_{\mathrm{bound}}=D$.
The corresponding $\mathbf{s}^{*}=[1.9116,1.9116]$ is kept fixed as $\mathbf{s}$ during the entire inversion process.

Samples of $p(\bm{\uptheta}|\mathbf{y})$ are generated from four Markov chains, with initial state of each chain drawn from the prior distribution $p(\bm{\uptheta})$. 
Each chain has $3000$ samples, with the first $1000$ samples discarded as burn-in. 
Sampling is performed using AdvancedHMC.jl \cite{Xu2020}, in which the leapfrog parameter $\epsilon$, and $L$, as well as the mass matrix $\mathbf{M}$, are automatically tuned during the burn-in period in a manner similar to Stan's HMC adaptation \cite{Carpenter2017}.

\begin{figure}[t]
\centering
\includegraphics{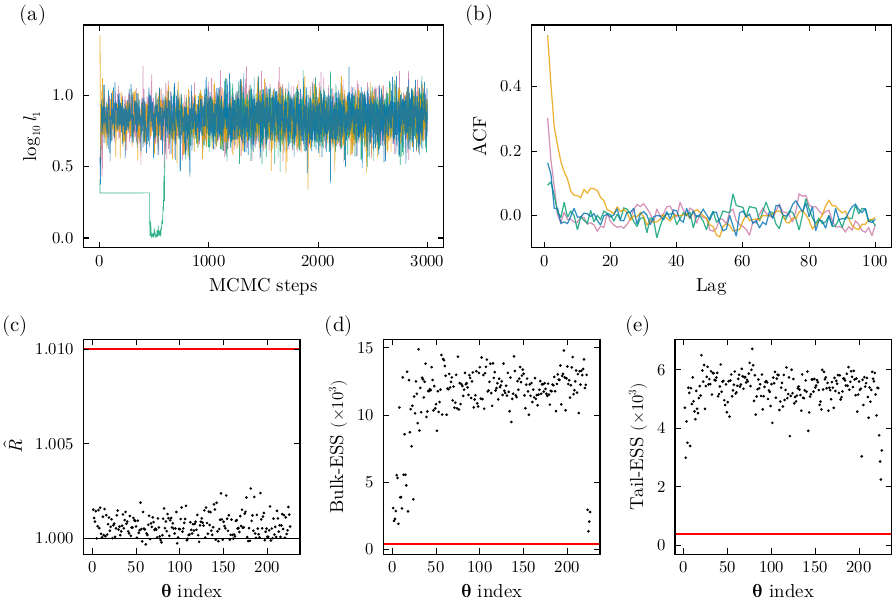}

\caption{Convergence Diagnostics for HMC. 
(a) Trace plots and (b) ACF plots for $\log_{10}l_{1}$ across all chains. 
(c) Rank-normalized $\hat{R}$, (d) Bulk-ESS, and (e) Tail-ESS for all elements of $\bm{\uptheta}$.
The red line in each panel indicates the thresholds, i.e., $\hat{R}=1.01$ and $\mathrm{ESS}=400$.}
\label{fig:test1_dx}
\end{figure}

As a representative sample, trace and autocorrelation function (ACF) plots for $\log_{10}l_{1}$ are shown in \cref{fig:test1_dx}(a) and \cref{fig:test1_dx}(b), respectively. 
All Markov chains converge in a similar range and exhibit well-mixed behavior.
The autocorrelations decay rapidly to near zero within a few lags, indicating efficient sampling and adequate convergence of the Markov chains. 
The convergence of Markov chains is also assessed using the $\hat{R}$ and the bulk/tail effective sample sizes (ESS), as implemented in ArviZ \cite{Kumar2019}. 
The $\hat{R}$ statistic measures chain convergence, while ESS quantifies the effective number of independent samples; bulk ESS reflects sampling efficiency in the central region of the posterior, whereas tail ESS evaluates it in the distribution tails \cite{Vehtari2021}. 
In this study, we define tail ESS as the minimum ESS associated with the $2.5$th and $97.5$th percentiles. 
The recommended thresholds are $\hat{R}<1.01$ and $\mathrm{ESS}>400$, respectively \cite{Vehtari2021}. 
As shown in \cref{fig:test1_dx}(c)--(e), all elements of $\bm{\uptheta}$ satisfy these criteria. 
This suggests adequate sampling quality across both the bulk and the tails of the posterior distribution.

\begin{figure}[t]
\centering
\includegraphics{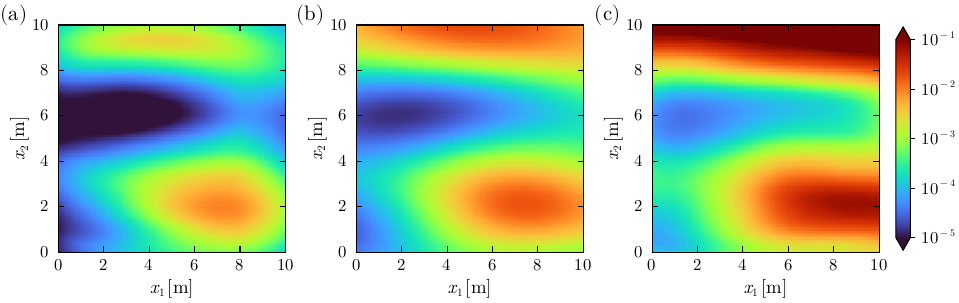}

\caption{Profiles of the posterior hydraulic conductivity $k|\mathbf{y}$: (a) $2.5$th, (b) $50$th (median), and (c) $97.5$th percentile fields.
Panels (a) and (c) correspond to the lower and upper bounds of the $95\%$ CI, respectively.}
\label{fig:test1_CI}
\end{figure}

\begin{figure}[t]
\centering
\includegraphics{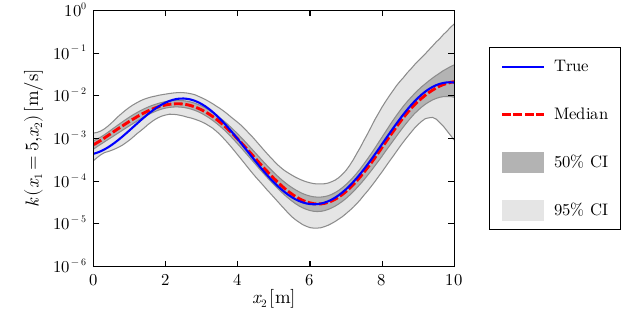}

\caption{Cross-sectional profiles of the posterior log-hydraulic conductivity field $u|\mathbf{y}$ at $x_{1}=5.$}
\label{fig:test1_cross_section}
\end{figure}

\begin{figure}[t]
\centering
\includegraphics{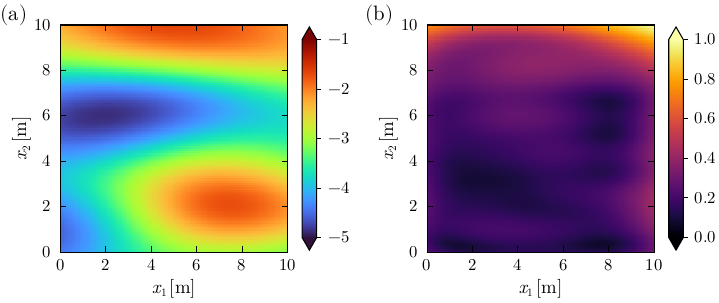}

\caption{The profiles of the posterior log-hydraulic conductivity field $u|\mathbf{y}$: (a) mean and (b) standard deviation.}
\label{fig:test1_mean_std}
\end{figure}

The profile of posterior hydraulic conductivity $k|\mathbf{y}$ is illustrated in \cref{fig:test1_CI}. 
The median field (\cref{fig:test1_CI}(b)) closely resembles the true field. 
The $95$\% credible interval (CI), bounded by the $2.5$th and $97.5$th percentile fields (\cref{fig:test1_CI}(a) and \cref{fig:test1_CI}(c)), contains the true field within an interval spanning approximately $1$--$2$ orders of magnitude. 
\cref{fig:test1_cross_section} shows a cross-sectional profile of $k|\mathbf{y}$ at $x_{1}=5$ as a representative example. 
In most regions, the true field aligns well with the median field. 
In the lower region ($x_{2}\leq5$), the true field is contained within a relatively narrow $95\%$ CI, indicating
low uncertainty. 
In the upper region ($x_{2}\geq5$), although the $95\%$ CI is wider than in the lower region, the $50\%$ CI ($25$th--$75$th percentile interval) still contains the true field. 
The behavior of uncertainty range is consistent with the statistics of the posterior log-hydraulic conductivity $u|\mathbf{y}$ shown in \cref{fig:test1_mean_std}.
The mean of $u|\mathbf{y}$ (\cref{fig:test1_mean_std}(a)) closely matches the true field. 
The standard deviation of $u|\mathbf{y}$ (\cref{fig:test1_mean_std}(b)) tends to be smaller in the lower region, which corresponds to a narrower uncertainty range. 
Although $u|\mathbf{y}$ is not strictly Gaussian, the uncertainty range can be evaluated as approximately $\pm2$ standard deviations from the mean. 
These results demonstrate that the posterior hydraulic conductivity captures the true field within a narrow uncertainty range, suggesting that the estimation is accurate.

\begin{figure}[tp]
\centering
\includegraphics{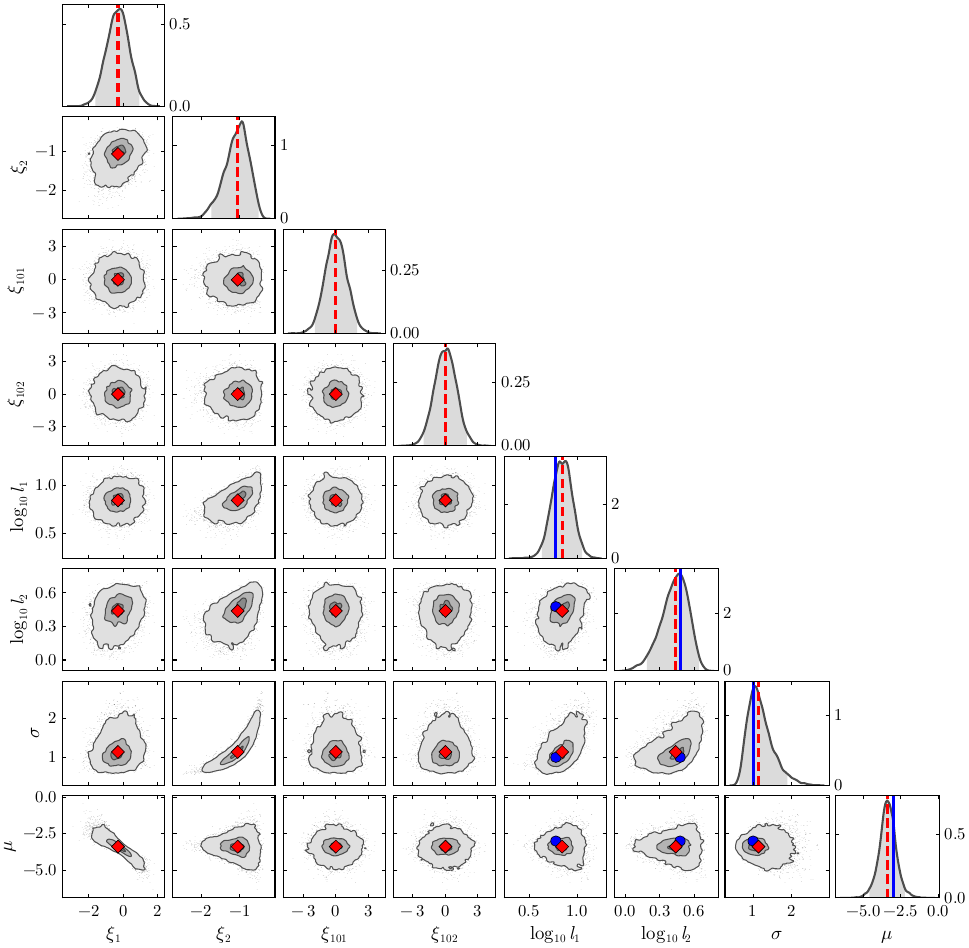}

\caption{Posterior marginal distributions of $(\xi_{1},\xi_{2},\xi_{101},\xi_{102},\log_{10}l_{1},\log_{10}l_{2},\sigma,\mu)$.
The red diamonds and dashed lines indicate the posterior medians.
For $(\log_{10}l_{1},\log_{10}l_{2},\sigma,\mu)$, blue circles and solid lines represent the true values. 
In the 1D distribution plots, the grey shaded areas indicate the $95\%$ HPD regions. 
In the 2D distribution plots, the three contours represent the $10\%$, $50\%$, and $95\%$ HPD regions.}
\label{fig:test1_jointplot}
\end{figure}

The posterior distributions of $(\xi_{1},\xi_{2},\xi_{101},\xi_{102},\log_{10}l_{1},\log_{10}l_{2},\sigma,\mu)$ are shown in \cref{fig:test1_jointplot}. 
For all hyperparameters $(\log_{10}l_{1},\log_{10}l_{2},\sigma,\mu)$, the $95\%$ Highest Probability Density (HPD) regions encompass the true values. 
Moreover, the posterior medians and the true values agree well, indicating accurate estimation of the hyperparameters
and effective capture of the characteristics of the random field.
Furthermore, the hyperparameters and KL coefficients corresponding to large eigenvalues dominantly contribute to the representation of the random field, and correlations are observed among some of these principal parameters in their posterior distributions (e.g., $\xi_{1}\text{--}\mu$,
$\xi_{2}\text{--}\sigma$). 
In contrast, KL coefficients associated with small eigenvalues (represented by $\xi_{101}$ and $\xi_{102}$) are almost independent of the other parameters, with each posterior distribution remaining close to the prior $\mathcal{N}(0,1)$. 
This is because these parameters have a negligible impact on the likelihood, causing the prior characteristics to remain dominant. 
Specifically, the KL expansion constructed to achieve $\overline{\varepsilon_{\sigma}}=10^{-2}$ for $l_{1}=l_{2}=1$ is overparameterized for representing the random fields with larger correlation lengths, such as those of the posterior median (corresponding to $l_{1}\approx6.9$ and $l_{2}\approx2.7$).
Consequently, the KL coefficients associated with small eigenvalues become redundant. 
This finding also indicates that the chosen lower bounds for the correlation lengths are sufficiently small to ensure
adequate representational capability of the KL expansion in the inversion process. 
Additionally, these redundant parameters exhibit a higher
ESS than the principal parameters (see \cref{fig:test1_dx}(d) and \cref{fig:test1_dx}(e)). 
This behavior reflects the relatively simple geometry of their posterior distributions, which stems from their prior distributions, namely independent normal distributions.

\begin{figure}[t]
\centering
\includegraphics{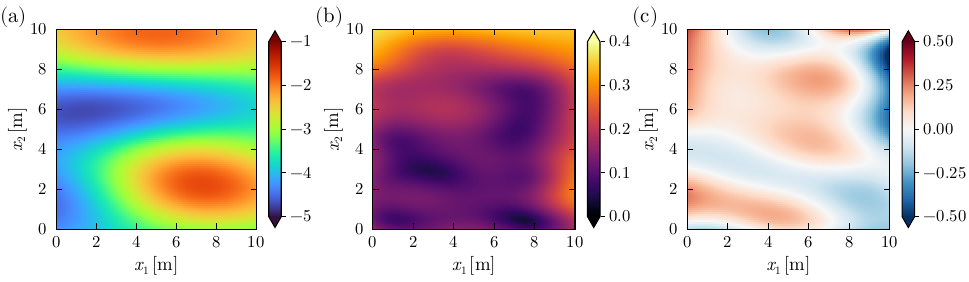}

\caption{(a) Mean, (b) standard deviation, and (c) bias of the posterior means $\mathrm{E}[u|\mathbf{y}_i]$ across the $100$ observational datasets.}
\label{fig:test1_robust}
\end{figure}

To verify that the above results are not incidental to a specific dataset, inverse analyses are performed on $100$ observational datasets $\{\mathbf{y}_i\}_{i=1}^{100}$ with identical noise characteristics. 
\cref{fig:test1_robust}(a) presents the mean of the posterior means across the 100 observational datasets, $(1/100)\sum_{i=1}^{100} \mathrm{E}[u|\mathbf{y}_i]$, where $\mathrm{E}[u|\mathbf{y}_i]$ denotes the posterior mean for the $i$-th dataset.
\cref{fig:test1_robust}(b) and \cref{fig:test1_robust}(c) show, respectively, the standard deviation of the posterior means and the bias (defined as the difference between the mean of the posterior means and the true field).
The standard deviation (\cref{fig:test1_robust}(b)) is within $0.4$ across the entire domain, indicating that the estimates do not significantly vary with differences in the observational data, which suggests a degree of robustness. 
Additionally, the mean (\cref{fig:test1_robust}(a)) agrees well with the true field (\cref{fig:test1_config}). 
Indeed, the bias has a small magnitude over most of the domain, as shown in \cref{fig:test1_robust}(c). 
These findings suggest that, under the current problem setting, estimates close to the true field can be achieved stably regardless of the observational data.

\subsection{Test 2}

\begin{figure}[t]
\centering
\includegraphics{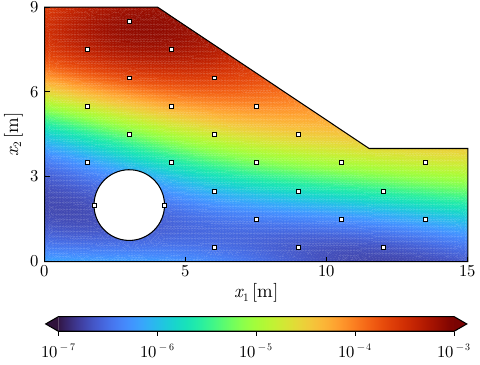}

\caption{The true hydraulic conductivity field generated through the log-hydraulic conductivity $u$ represented by the KL expansion with $(l_{1},l_{2},\sigma,\mu)=(10,5,1,-5)$.
White squares indicate observation points for the hydraulic head.}
\label{fig:test2_config}
\end{figure}

Consider the hydraulic conductivity field $k$ over the domain $D$ with a complex geometry, as shown in \cref{fig:test2_config}. 
The domain $D$ includes a circle centered at $(x_{1},x_{2})=(3,2)$ with a radius of $1.25$. 
The true field is generated numerically through the log-hydraulic conductivity $u$, represented by the KL expansion with hyperparameters $(l_{1},l_{2},\sigma,\mu)=(10,5,1,-5)$. 
The source term is $Q=0$, and boundary conditions are defined as
\begin{equation}
\begin{aligned}
h(\mathbf{x})=9\quad\text{on }\Gamma_\mathrm{left}, & \quad h(\mathbf{x})=4\quad\text{on }\Gamma_\mathrm{right},\\
h(\mathbf{x})=x_{2}\quad\text{on }\Gamma_\mathrm{top}, & \quad\frac{\partial h}{\partial\mathbf{n}}(\mathbf{x})=0\quad\text{on }\Gamma_{\mathrm{N}},
\end{aligned}
\label{eq:bc_test2}
\end{equation}
where $\Gamma_\mathrm{left}=\{\mathbf{x}\in \partial D \mid x_1 = 0\}$, $\Gamma_\mathrm{right}=\{\mathbf{x}\in \partial D \mid x_1 = 15\}$, $\Gamma_\mathrm{top}=\Gamma_\mathrm{D}\backslash(\Gamma_\mathrm{left}\cup\Gamma_\mathrm{right})$, and $\Gamma_{\mathrm{N}}=\{\mathbf{x}\in\partial D\mid x_{2}=0\}\cup\{\mathbf{x}\in\partial D\mid(x_{1}-3)^{2}+(x_{2}-2)^{2}=1.25^{2}\}$.
These boundary conditions describe gravity-driven seepage flow in saturated soil and can be interpreted as representing subsurface groundwater flow within a slope that becomes fully saturated due to heavy rainfall.
The governing equation is homogeneous (\cref{eq:PDE} with $Q=0$), and homogeneous Neumann boundary conditions are imposed.
Consequently, for a given hydraulic head $h$, the hydraulic conductivity $k$ is determined only up to a positive multiplicative constant.
Therefore, the absolute scale of $k$ cannot be uniquely determined from observations of $h$ alone, and flux information is essential for its identification. 
Accordingly, in this test case, the observational data consist of the total hydraulic head at $27$ points in $D$ (white squares in \cref{fig:test2_config}), together with the flow rate from the right edge of $D$, i.e., $\{\mathbf{x}\in\partial D|x_{1}=15\}$.
The noise-free observations are synthetically generated using FEM with finer FE meshes than those used for inversion.
Assuming the observation points are known precisely, the elevation heads are error-free; therefore, all observation errors in the total head originate solely from the pressure head. 
Consequently, independent noise is added to both the pressure head and the flow rate, with a standard deviation of $1.5\%$ of their respective values.

The prior for $k$ is set to the log-normal random field, as in Test 1. 
The prior for the random coefficients $\bm{\upxi}$ in the KL expansion is the standard normal $\bm{\upxi}\sim\mathcal{N}(\mathbf{0},\mathbf{I})$, while hyperparameters are assigned weakly informative priors: 
$\log_{10}(l_{1}/l^{\min}_{1})\sim\mathcal{HN}(0,1)$, $\log_{10}(l_{2}/l^{\min}_{2})\sim\mathcal{HN}(0,1)$, $\sigma\sim\mathcal{HN}(0,1)$, and $\mu\sim\mathcal{N}(-4,2)$.
The lower bounds are set to $l^{\min}_{1}=1.5$ and $l^{\min}_{2}=1.0$, respectively, which correspond to approximately one-tenth of the lengths of the bounding domain $D_{\mathrm{bound}}=[0,15]\times[0,9]$ in each spatial direction.
As in Test 1, the number of the KL expansion terms $M$ is set to $M^{*}=200$, which is obtained by solving \cref{prob:2} on $D_{\mathrm{bound}}$ with $\overline{\varepsilon_{\sigma}}=10^{-2}$, $l_{1}=l^{\min}_{1}$, and $l_{2}=l^{\min}_{2}$. The corresponding $\mathbf{s}^{*}=[2.8401,1.6813]$ is kept fixed as $\mathbf{s}$ during the inversion process.

\begin{figure}[t]
\centering
\includegraphics{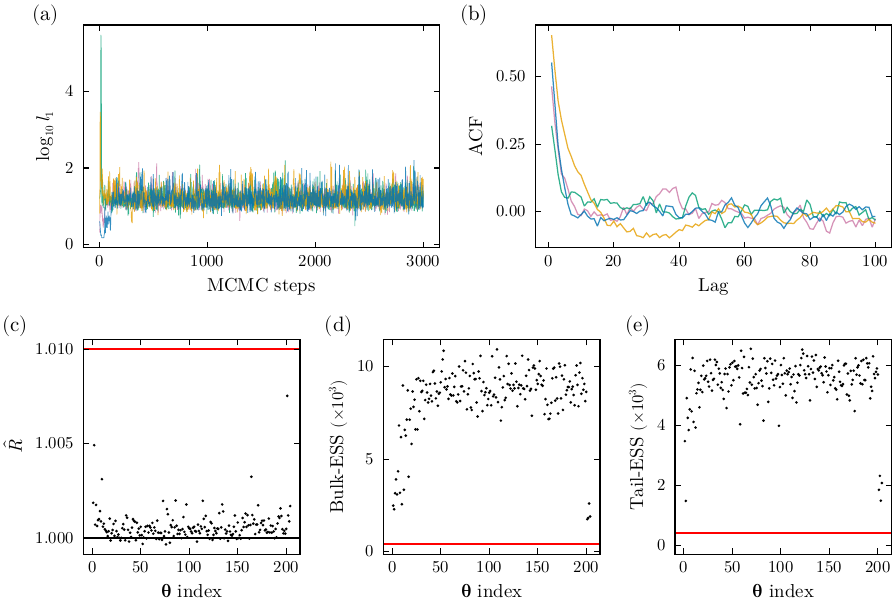}

\caption{Convergence Diagnostics for HMC. 
(a) Trace plots and (b) ACF plots for $\theta_{1}$ across all chains. 
(c) Rank-normalized $\hat{R}$, (d) Bulk-ESS, and (e) Tail-ESS for all elements of $\bm{\uptheta}$.
The red line in each panel indicates the thresholds, i.e., $\hat{R}=1.01$ and $\mathrm{ESS}=400$.}\label{fig:test2_dx}
\end{figure}

The HMC sampling settings are the same as in Test 1: 
Using AdvancedHMC.jl, four Markov chains are run with initial states drawn from the prior distribution. 
Each chain consists of 3000 samples, with the first
$1000$ discarded as burn-in. 
The HMC diagnostic results are shown in \cref{fig:test2_dx}. 
Representative trace and ACF plots for $\log_{10}l_{1}$ are shown in \cref{fig:test2_dx}(a) and \cref{fig:test2_dx}(b), respectively, indicating sufficient convergence and efficient sampling of Markov chains. 
All elements of $\bm{\uptheta}$ satisfy the recommended thresholds for the rank-normalized $\hat{R}$ and the bulk/tail ESS as shown in \cref{fig:test2_dx}(c)--(e); specifically, $\hat{R}<1.01$ and $\mathrm{ESS}>400$. 
These results suggest adequate sampling quality across both the bulk and the tails of the posterior distribution.

\begin{figure}[tp]
\centering
\includegraphics{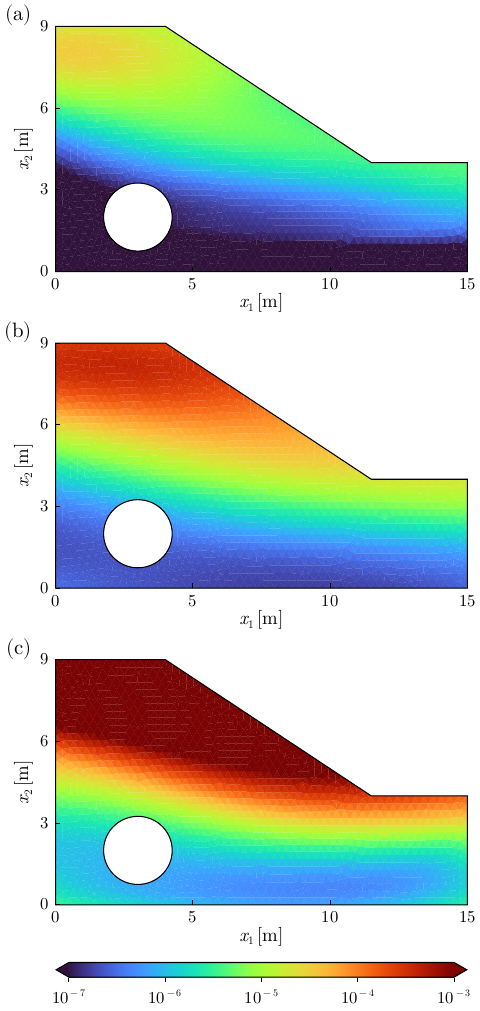}

\caption{The profiles of the posterior hydraulic conductivity $k|\mathbf{y}$: (a) $2.5$th, (b) $50$th (median), and (c) $97.5$th percentile fields.
Panels (a) and (c) correspond to the lower and upper bounds of the $95\%$ CI, respectively.}
\label{fig:test2_CI}
\end{figure}

\begin{figure}[t]
\centering
\includegraphics{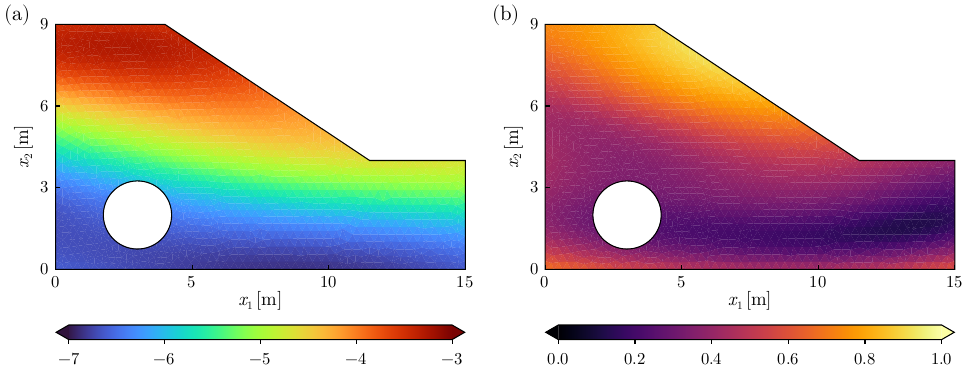}

\caption{The profiles of the posterior log-hydraulic conductivity field $u|\mathbf{y}$:
(a) mean and (b) standard deviation.}
\label{fig:test2_mean_std}
\end{figure}

The $2.5$th, $50$th, and $97.5$th percentile fields of the posterior hydraulic conductivity $k|\mathbf{y}$ are shown in \cref{fig:test2_CI}.
The $50$th percentile (median) field (\cref{fig:test2_CI}(b)) closely resembles the true field, and \cref{fig:test2_CI}(a) and \cref{fig:test2_CI}(c) show that the $95\%$ CI contains the true field. 
However, the width of the $95\%$ CI is larger near the top of the domain $D$, indicating a relatively high degree of uncertainty. 
A similar pattern in uncertainty can be observed in the mean and standard deviation of the log-hydraulic conductivity $u|\mathbf{y}$, as shown in \cref{fig:test2_mean_std}.
Although the mean closely resembles the true field, the standard deviation of $u|\mathbf{y}$ tends to be higher in the upper part of $D$. 
This trend in uncertainty can be attributed to the boundary conditions in \cref{eq:bc_test2}. 
Since the upper part of $D$ is surrounded by the Dirichlet boundaries, the hydraulic head does not change significantly
even when hydraulic conductivity varies. 
This makes it difficult to identify the hydraulic conductivity within this region. 
Conversely, in the lower part of $D$, which is relatively distant from the Dirichlet boundary, variations in hydraulic conductivity are reflected in the observed hydraulic heads. 
Consequently, the posterior estimates of the hydraulic conductivity in this region exhibit lower uncertainty.

\begin{figure}[tp]
\centering
\includegraphics{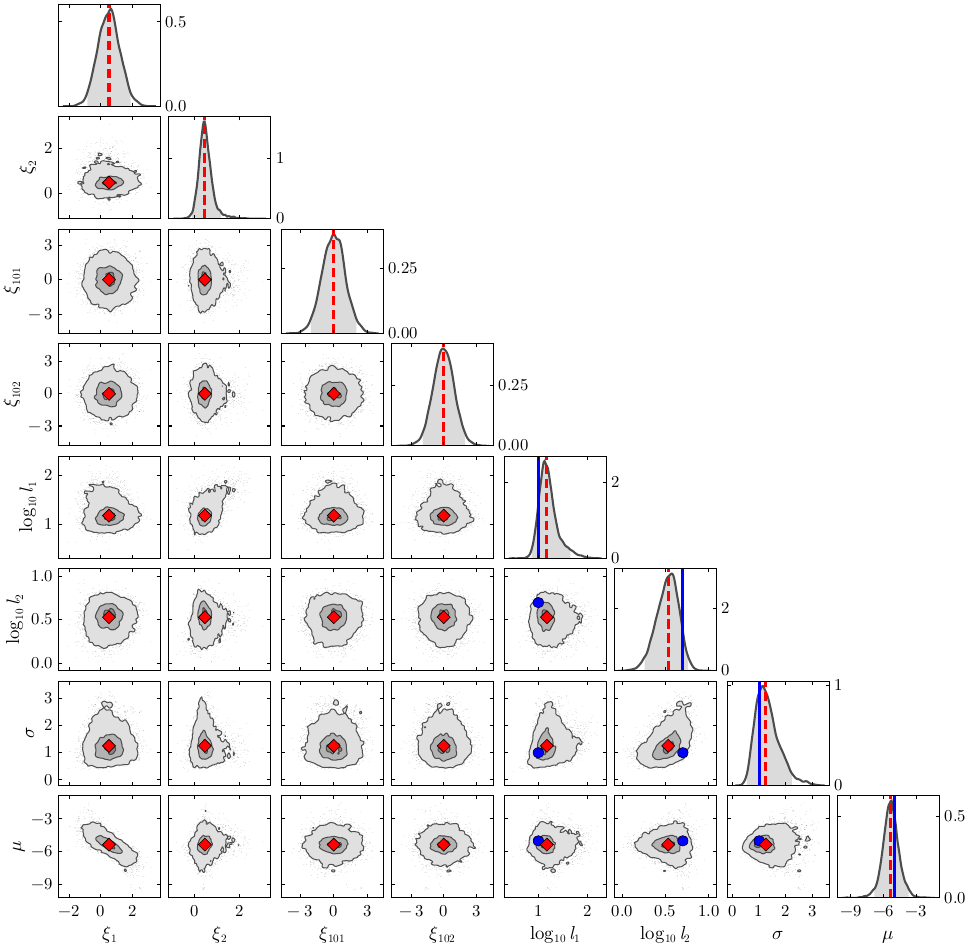}

\caption{Posterior marginal distributions of $(\xi_{1},\xi_{2},\xi_{101},\xi_{102},\log_{10}l_{1},\log_{10}l_{2},\sigma,\mu)$.
The red diamonds and dashed lines indicate the posterior medians.
For $(\log_{10}l_{1},\log_{10}l_{2},\sigma,\mu)$, blue circles and solid lines represent the true values. 
In the 1D distribution plots, the grey shaded areas indicate the $95\%$ HPD regions. 
In the 2D distribution plots, the three contours represent the $10\%$, $50\%$, and $95\%$ HPD regions.}
\label{fig:test2_jointplot}
\end{figure}

\begin{figure}[t]
\centering
\includegraphics{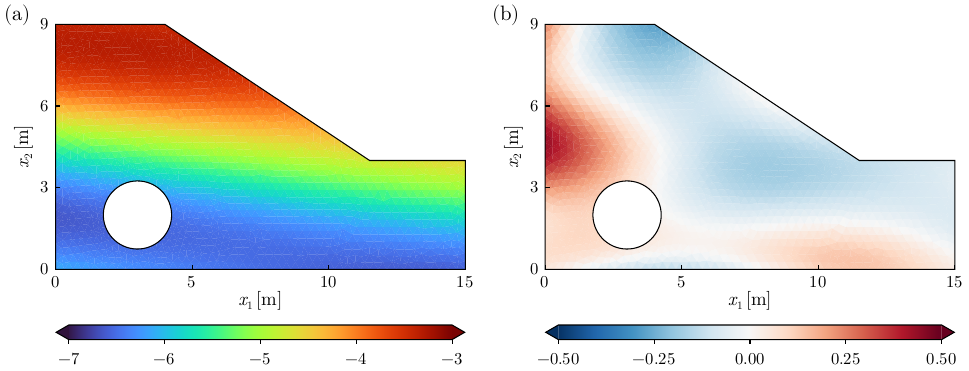}

\caption{(a) A sample from the posterior $u|\mathbf{y}$, with $(l_{1},l_{2},\sigma,\mu)=(23.795,3.4590,0.90983,-4.5215)$. 
(b) The residual relative to the true field with $(l_{1},l_{2},\sigma,\mu)=(10,5,1,-5)$, whose magnitude is less than $0.2$ over most of the domain and is relatively small.}\label{fig:test2_post_sample}
\end{figure}

The posterior distributions of $(\xi_{1},\xi_{2},\xi_{101},\xi_{102},\log_{10}l_{1},\log_{10}l_{2},\sigma,\mu)$ are shown in \cref{fig:test2_jointplot}. 
For all hyperparameters $(\log_{10}l_{1},\log_{10}l_{2},\sigma,\mu)$, the $95\%$ HPD regions encompass the true values. 
For the correlation lengths, the true values deviate from the posterior medians and lie near the edges of the $95\%$ HPD region, indicating that the estimation accuracy may be somewhat lower. 
However, this is not a critical issue because the main focus is the estimation of the hydraulic conductivity field, while the hyperparameter estimates themselves serve only as auxiliary indicators. 
\cref{fig:test2_post_sample} shows a sample of $u|\mathbf{y}$, generated with $(l_{1},l_{2},\sigma,\mu)=(23.795,3.4590,0.90983,-4.5215)$, along with the residual compared to the true field. 
These hyperparameters differ from $(l_{1},l_{2},\sigma,\mu)=(10,5,1,-5)$, which are used to generate the true field, particularly in the correlation lengths.
Nevertheless, the two fields exhibit similar spatial structures. 
This indicates that similar fields can be generated even with different hyperparameter settings. 
In the presence of such non-identifiability, fixing the hyperparameters to a single value (e.g., the maximum likelihood estimation) may not be sufficient. 
This insight highlights the importance of adopting a hierarchical Bayesian approach, where hyperparameters are treated as random variables to properly quantify uncertainty.
For the KL coefficients, similar behavior is observed in Test 1. 
Specifically, the posterior distributions of $\xi_{1}$ and $\xi_{2}$, associated with the larger eigenvalues, show correlations with several parameters (e.g., $\xi_{1}\text{--}\mu$, $\xi_{2}\text{--}\log_{10}l_{1}$).
In contrast, the posterior distributions of $\xi_{101}$ and $\xi_{102}$, corresponding to small eigenvalues, closely resemble their prior distributions $\mathcal{N}(0,1)$, indicating that they are almost independent of the other parameters.

\begin{figure}[t]
\centering
\includegraphics{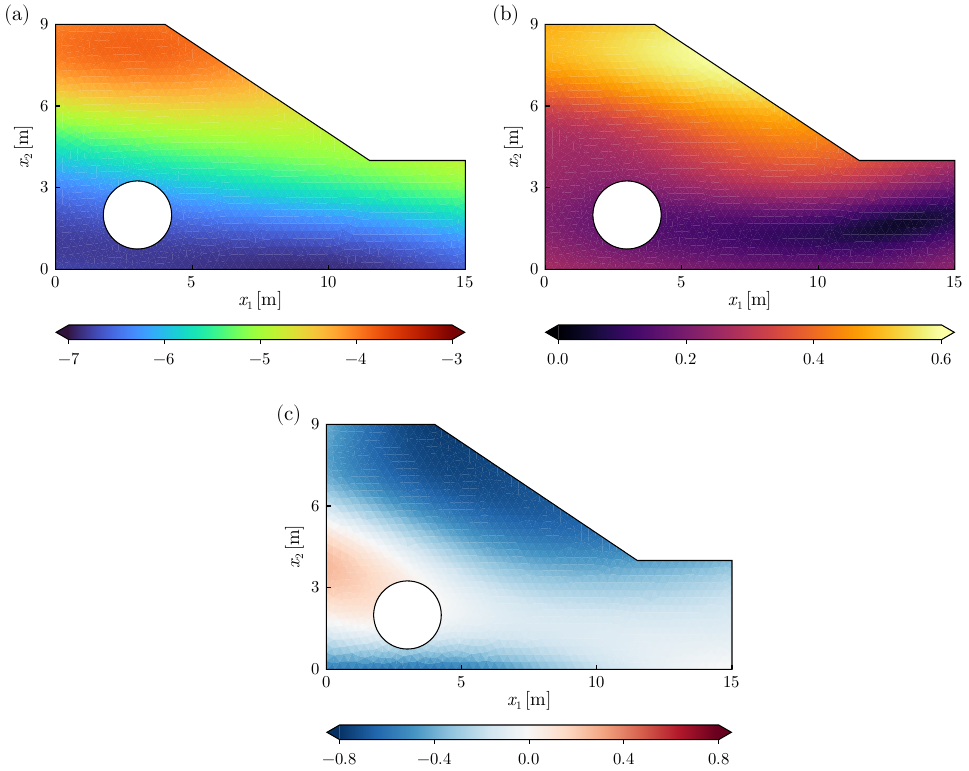}

\caption{(a) Mean, (b) standard deviation, and (c) bias of the posterior means $\mathrm{E}[u|\mathbf{y}_i]$ across the $100$ observational datasets.}\label{fig:test2_robust}
\end{figure}

As in Test 1, inverse analyses are performed on $100$ observational datasets $\{\mathbf{y}_i\}_{i=1}^{100}$ with identical noise characteristics. 
\cref{fig:test2_robust} shows the mean, standard deviation, and bias of the posterior means $\mathrm{E}[u|\mathbf{y}_i]$ ($i=1,\ldots,100$).
Although a notable negative bias and higher standard deviation are observed in the upper region, the overall trend of the mean is generally consistent with the true field, particularly in the lower region. 
These results suggest that the estimation achieves a certain level of accuracy regardless of the observational data, indicating that the results for the representative dataset are not merely coincidental.
It should be noted that the reduced accuracy in the upper region is likely attributable to the Dirichlet boundary condition, as discussed in the representative case.

\section{Conclusions}

We modeled a Gaussian random field with a squared exponential kernel using the Karhunen--Loève (KL) expansion based on the analytical solution to an integral eigenvalue problem (IEVP), with the aim of applying it to hierarchical Bayesian inversion. 
The analytical solution of the squared exponential kernel exists for a Gaussian-weighted IEVP defined on $\mathbb{R}$. 
Although the KL expansion based on the analytical eigenpairs (analytical KL expansion) requires a larger number of terms than the conventional KL expansion to achieve a comparable level of approximation accuracy, it retains the domain independence property and enables the generation of Gaussian random field samples over any closed physical domain.

The standard deviation of the Gaussian weight function can be chosen as any positive value.
In this study, we propose an optimization-based approach for selecting both the standard deviation and the number of KL terms to achieve high approximation accuracy with as few KL terms as possible. Specifically, optimal parameters are determined by minimizing the mean error variance. 
To simplify the optimization process, the mean error variance is evaluated over the hyperrectangle bounding domain rather than the physical domain. 
Consequently, it is confirmed that the mean error variance in the physical domain is approximately comparable to the error tolerance prescribed for the bounding domain.

By employing this analytical solution, repeated IEVP computations during posterior sampling in hierarchical Bayesian inversion can be avoided, thereby eliminating a computational bottleneck. 
Furthermore, the analytical KL expansion can be differentiated in closed-form with respect to all parameters, including the KL coefficients and hyperparameters.
This property enables efficient posterior sampling via Hamiltonian Monte Carlo (HMC).

Finally, we applied the analytical KL expansion to a Bayesian inverse problem to estimate the hydraulic conductivity field in two-dimensional Darcy flow. 
A lower bound for the correlation lengths is prescribed
to represent sufficiently high-frequency spatial variations. 
The standard deviation of the Gaussian weight function and the number of KL terms are then determined to meet the prescribed error tolerance, even under this most demanding condition. 
These values are fixed throughout the entire inversion process. 
The posterior random field is successfully characterized
in terms of typical magnitude, fluctuation amplitude, and spatial frequency under weakly informative priors on hyperparameters. 
These priors are chosen to adequately represent hydraulic conductivity relevant to engineering applications. 
Furthermore, it is shown that the analytical KL expansion is applicable to inverse problems not only on a square domain but also on a domain with complex geometry. 
This result suggests a potential extension in which the analytical KL expansion can be utilized as a component for constructing complex random fields. 
For instance, by applying it to each zoning subdomain separately \cite{Tipireddy2017,Tipireddy2018,Xu2024}, it is
possible to represent non-stationary fields with distinct statistical properties across different regions. 
This approach effectively introduces discontinuities and local structures, thereby alleviating the smoothness of the squared exponential kernel.

\section*{Acknowledgements}
This work was supported by JSPS KAKENHI Grant Numbers JP25KJ1594 and JP25H00951.

\appendix
\crefalias{section}{appendix}

\section{Adjoint method}\label{app:adjoint_method}

The adjoint method provides an efficient way to compute the gradient of the potential energy $\partial U(\bm{\uptheta})/\partial\bm{\uptheta}$ without explicitly evaluating the sensitivities of the forward problem with respect to each parameter in $\bm{\uptheta}$. 
Let $\mathcal{G}_{h,n}$ and $\mathcal{G}_{q,n}$ denote the $n$-th components of $\mathcal{G}_{h}$ and $\mathcal{G}_{q}$, respectively, defined as 
\begin{equation}
\mathcal{G}_{h,n}(\bm{\uptheta})=\left(O_{h,n}\circ G_{h}\right)(k(\cdot;\bm{\uptheta})),\quad\mathcal{G}_{q,n}(\bm{\uptheta})=\left(O_{q,n}\circ G_{q}\right)(k(\cdot;\bm{\uptheta})).
\end{equation}
According to \cref{eq:dg_dtheta2,eq:obs_h,eq:obs_q}, the derivatives of $\mathcal{G}_{h,n}$ and $\mathcal{G}_{q,n}$ with respect to $\theta_{j}$ are given by
\begin{equation}
\frac{\partial\mathcal{G}_{h,n}}{\partial\theta_{j}}(\bm{\uptheta})=\int_{D}\frac{\partial h}{\partial\theta_{j}}(\mathbf{x};\bm{\uptheta})\delta(\mathbf{x}-\mathbf{x}_{n})\,\mathrm{d}\mathbf{x},\label{eq:deriv_obs_h}
\end{equation}
\begin{equation}
\frac{\partial\mathcal{G}_{q,n}}{\partial\theta_{j}}(\bm{\uptheta})=\int_{\Gamma_{q,n}}\frac{\partial\mathbf{q}}{\partial\theta_{j}}(\mathbf{x};\bm{\uptheta})\cdot\mathbf{n}\,\mathrm{d}S(\mathbf{x}).\label{eq:deriv_obs_q}
\end{equation}
These expressions explicitly depend on the forward sensitivities $\partial h/\partial\theta_{j}$ and $\partial\mathbf{q}/\partial\theta_{j}$, the computation of which for all $\theta_{j}$ become prohibitively expensive in high-dimensional parameter spaces. 
To avoid this difficulty, an adjoint problem is introduced.

Let $\varphi$ be an adjoint field satisfying the adjoint PDE
\begin{equation}
-\nabla\cdot\left(k\nabla\varphi\right)=\sum^{N_{h}}_{n=1}\widetilde{Q}_{n}\delta(\mathbf{x}-\mathbf{x}_{n})\quad\text{in }D,\label{eq:adjoint_PDE}
\end{equation}
subject to the boundary conditions:
\begin{equation}
\varphi=\begin{cases}
0 & \text{on }\Gamma_{\mathrm{D}}\backslash\left(\cup^{N_{q}}_{n=1}\Gamma_{q,n}\right)\\
\widetilde{\varphi}_{n} & \text{on }\Gamma_{q,n}
\end{cases},\quad\nabla\varphi\cdot\mathbf{n}=0\quad\text{on }\Gamma_{\mathrm{N}}.
\end{equation}
Here, the adjoint source terms $\widetilde{Q}_{n}$ and the adjoint boundary values $\widetilde{\varphi}_{n}$ are the $n$-th components of the vectors $\widetilde{\mathbf{Q}}$ and $\widetilde{\bm{\upvarphi}}$, respectively, which are defined as
\begin{equation}
\widetilde{\mathbf{Q}}=-\mathbf{R}^{-1}_{h}\left(\mathbf{y}_{h}-\mathcal{G}_{h}(\bm{\uptheta})\right),\quad\widetilde{\bm{\upvarphi}}=-\mathbf{R}^{-1}_{q}\left(\mathbf{y}_{q}-\mathcal{G}_{q}(\bm{\uptheta})\right).
\end{equation}
These adjoint quantities are designed such that the forward sensitivity terms in $\partial U(\bm{\uptheta})/\partial\bm{\uptheta}$ can be eliminated, as demonstrated below.

Differentiating the original problem \cref{eq:PDE,eq:bc} with respect to $\theta_{j}$ yields
\begin{equation}
\frac{\partial}{\partial\theta_{j}}\left(-\nabla\cdot\left(k\nabla h\right)\right)=\nabla\cdot\frac{\partial\mathbf{q}}{\partial\theta_{j}}=0\quad\text{in }D,\label{eq:deriv_PDE}
\end{equation}
subject to the boundary conditions
\begin{equation}
\frac{\partial h}{\partial\theta_{j}}=0\quad\mathrm{on}\ \Gamma_{\mathrm{D}},\quad\frac{\partial\mathbf{q}}{\partial\theta_{j}}\cdot\mathbf{n}=0\quad\text{on }\Gamma_{\mathrm{N}},
\end{equation}
where invariance of $Q$, $h_{\mathrm{D}}$, and $q_{\mathrm{N}}$ with respect to $\theta_{j}$ has been used. 
Hereafter, the dependence of $k$, $h$, and $\mathbf{q}$ on $\bm{\uptheta}$ is omitted for notational simplicity.

Define the Lagrangian by incorporating the PDE constraint as
\begin{equation}
\begin{aligned}\mathcal{L}(\bm{\uptheta}) & =\frac{1}{2}(\mathbf{y}-\mathcal{G}(\bm{\uptheta}))^{\top}\mathbf{R}^{-1}(\mathbf{y}-\mathcal{G}(\bm{\uptheta}))+\int_{D}\varphi\,\Bigl(\nabla\cdot(k\nabla h)+Q\Bigr)\,\mathrm{d}\mathbf{x}\\
 & =\frac{1}{2}(\mathbf{y}-\mathcal{G}(\bm{\uptheta}))^{\top}\mathbf{R}^{-1}(\mathbf{y}-\mathcal{G}(\bm{\uptheta}))+\int_{D}\varphi\,\Bigl(-\nabla\cdot\mathbf{q}+Q\Bigr)\,\mathrm{d}\mathbf{x}.
\end{aligned}
\end{equation}
Since the second term is identically zero due to the PDE constraint, we obtain
\begin{equation}
U(\bm{\uptheta}) = \mathcal{L}(\bm{\uptheta}) + E(\bm{\uptheta}).\label{eq:L_E}
\end{equation}
Differentiating \cref{eq:L_E} with respect to $\theta_j$ yields
\begin{equation}
\begin{aligned}
\frac{\partial U}{\partial\theta_{j}}(\bm{\uptheta})&=-(\mathbf{y}-\mathcal{G}(\bm{\uptheta}))^{\top}\mathbf{R}^{-1}\frac{\partial\mathcal{G}}{\partial\theta_{j}}(\bm{\uptheta})-\int_{D}\varphi\nabla\cdot\frac{\partial\mathbf{q}}{\partial\theta_{j}}\,\mathrm{d}\mathbf{x}+\frac{\partial E}{\partial\theta_{j}}(\bm{\uptheta})\\&=\widetilde{\mathbf{Q}}^{\top}\frac{\partial\mathcal{G}_{h}}{\partial\theta_{j}}(\bm{\uptheta})+\widetilde{\bm{\upvarphi}}^{\top}\frac{\partial\mathcal{G}_{q}}{\partial\theta_{j}}(\bm{\uptheta})-\int_{D}\varphi\nabla\cdot\frac{\partial\mathbf{q}}{\partial\theta_{j}}\,\mathrm{d}\mathbf{x}+\frac{\partial E}{\partial\theta_{j}}(\bm{\uptheta}),
\end{aligned}\label{eq:dUdtheta}
\end{equation}
where the volume integral term can be rewritten using integration by parts as follows:
\begin{equation}
-\int_{D}\varphi\nabla\cdot\frac{\partial\mathbf{q}}{\partial\theta_{j}}\,\mathrm{d}\mathbf{x}=-\int_{\partial D}\varphi\left(\frac{\partial\mathbf{q}}{\partial\theta_{j}}\cdot\mathbf{n}\right)\mathrm{d}S(\mathbf{x})+\int_{D}\nabla\varphi\cdot\frac{\partial\mathbf{q}}{\partial\theta_{j}}\,\mathrm{d}\mathbf{x}.\label{eq:adjoint_term1}
\end{equation}
Using the boundary conditions on $\partial\mathbf{q}/\partial\theta_{j}$ and $\varphi$, the boundary integral reduces to
\begin{equation}
-\int_{\partial D}\varphi\left(\frac{\partial\mathbf{q}}{\partial\theta_{j}}\cdot\mathbf{n}\right)\mathrm{d}S(\mathbf{x})=-\sum^{N_{q}}_{n=1}\widetilde{\varphi}_{n}\int_{\Gamma_{q,n}}\frac{\partial\mathbf{q}}{\partial\theta_{j}}\cdot\mathbf{n}\,\mathrm{d}S(\mathbf{x})=-\widetilde{\bm{\upvarphi}}^{\top}\frac{\partial\mathcal{G}_{q}}{\partial\theta_{j}}(\bm{\uptheta}).
\end{equation}
Recalling Darcy's law $\mathbf{q}=-k\nabla h$, the remaining volume integral becomes
\begin{equation}
\int_{D}\nabla\varphi\cdot\frac{\partial\mathbf{q}}{\partial\theta_{j}}\,\mathrm{d}\mathbf{x}=-\int_{D}k\left(\nabla\varphi\cdot\nabla\frac{\partial h}{\partial\theta_{j}}\right)\mathrm{d}\mathbf{x}-\int_{D}\frac{\partial k}{\partial\theta_{j}}\left(\nabla\varphi\cdot\nabla h\right)\mathrm{d}\mathbf{x}.
\end{equation}
The first term is further integrated by parts:
\begin{equation}
-\int_{D}k\left(\nabla\varphi\cdot\nabla\frac{\partial h}{\partial\theta_{j}}\right)\mathrm{d}\mathbf{x}=-\int_{\partial D}\frac{\partial h}{\partial\theta_{j}}k\left(\nabla\varphi\cdot\mathbf{n}\right)\mathrm{d}S(\mathbf{x})+\int_{D}\frac{\partial h}{\partial\theta_{j}}\nabla\cdot\left(k\nabla\varphi\right)\mathrm{d}\mathbf{x}.
\end{equation}
The boundary integral term vanishes due to the boundary conditions on $\partial h/\partial\theta_{j}$ and $\varphi$. 
Substituting the adjoint PDE \cref{eq:adjoint_PDE} into the remaining volume integral yields
\begin{equation}
\int_{D}\frac{\partial h}{\partial\theta_{j}}\nabla\cdot\left(k\nabla\varphi\right)\mathrm{d}\mathbf{x}=-\sum^{N_{h}}_{n=1}\widetilde{Q}_{n}\int_{D}\frac{\partial h}{\partial\theta_{j}}\delta(\mathbf{x}-\mathbf{x}_{n})\,\mathrm{d}\mathbf{x}=-\widetilde{\mathbf{Q}}^{\top}\frac{\partial\mathcal{G}_{h}}{\partial\theta_{j}}(\bm{\uptheta}).
\end{equation}
Combining the above results, \cref{eq:dUdtheta} is rewritten as
\begin{equation}
\frac{\partial U}{\partial\theta_{j}}(\bm{\uptheta})=-\int_{D}\frac{\partial k}{\partial\theta_{j}}\left(\nabla\varphi\cdot\nabla h\right)\mathrm{d}\mathbf{x}+\frac{\partial E}{\partial\theta_{j}}(\bm{\uptheta}).
\end{equation}
Hence, the term involving $\partial\mathcal{G}(\bm{\uptheta})/\partial\theta_{j}$, which requires expensive forward sensitivity computation, is eliminated and replaced by an expression involving the adjoint field $\varphi$.
As a result, the computation of the forward sensitivities for all parameters $\bm{\uptheta}$ is reduced to a single adjoint solve.

\bibliographystyle{elsarticle-num} 
\bibliography{reference}

\end{document}